\documentclass[3p, twocolumn]{elsarticle}
\usepackage{amsmath}
\usepackage{physics}
\usepackage{amsthm}
\usepackage{amsfonts}
\usepackage{bm}
\usepackage{graphicx}
\usepackage{booktabs}
\usepackage{tabularx}
\usepackage{longtable}
\usepackage{xcolor}
\usepackage{float}
\usepackage{url}
\usepackage[shortlabels]{enumitem}
\usepackage{soul}
\usepackage{amssymb}
\usepackage{hyperref}
\usepackage[normalem]{ulem}
\usepackage{comment}
\usepackage{epstopdf}
\usepackage{adjustbox}
\usepackage{xspace}
\usepackage{centernot}
\usepackage{multirow}
\usepackage{makecell}

\usepackage[skins]{tcolorbox}
\usepackage{subcaption}
\usetikzlibrary{shapes.geometric, arrows, calc, fit, positioning, chains, arrows.meta}

\newcommand{\mypar}[1]{\smallskip\noindent\textbf{#1.}\xspace}
\newcommand{\myparb}[1]{\smallskip\noindent\textbf{#1:}\xspace}

\newtheorem{theorem}{Theorem}[section]

\newtheorem{definition}[theorem]{Definition}

\newcommand{\C}{\mathbb{C}}
\newcommand{\R}{\mathbb{R}}
\newcommand{\N}{\mathbb{N}}
\newcommand{\Ot}{\widetilde{O}}
\newcommand{\thh}{\text{th}}
\newcommand{\quotes}[1]{``#1''}
\newcommand{\stress}[1]{\emph{#1}}

\newcommand{\m}[1]{{\mathbf{#1}}}
\newcommand{\ve}[1]{{\mathbf{#1}}}

\title{Evaluating the Potential of Quantum Machine Learning in Cybersecurity:\\
A Case-Study on PCA-based Intrusion Detection Systems}

\author[1]{Armando Bellante\fnref{fn1}\corref{cor1}}
\author[1]{Tommaso Fioravanti\fnref{fn2}}
\author[1]{Michele Carminati\fnref{fn3}}
\author[1]{Stefano Zanero\fnref{fn4}}
\author[2,3]{Alessandro Luongo\fnref{fn5}\corref{cor1}}

\affiliation[1]{organization={Politecnico di Milano}, 
addressline={Via Ponzio 34/5},
postcode={20133}, 
city={Milano}, 
country={Italia}}
\affiliation[2]{organization={Centre for Quantum Technologies}, 
addressline={Block S15, 3 Science Drive 2},
city={Singapore},
postcode={117543}, 
country={Singapore}}
\affiliation[3]{organization={Inveriant Pte. Ltd.},
addressline={101 Cecil Street \#14-12}, 
city={Singapore}, 
postcode={069533}, 
country={Singapore}}

\cortext[cor1]{Corresponding author}
\fntext[fn1]{\href{mailto:armando.bellante@polimi.it}{armando.bellante@polimi.it}}
\fntext[fn2]{\href{mailto:tommaso.fioravanti@mail.polimi.it}{tommaso.fioravanti@mail.polimi.it}}
\fntext[fn3]{\href{mailto:michele.carimati@polimi.it}{michele.carimati@polimi.it}}
\fntext[fn4]{\href{mailto:stefano.zanero@polimi.it}{stefano.zanero@polimi.it}}
\fntext[fn5]{\href{mailto:ale@nus.edu.sg}{ale@nus.edu.sg}}

\begin{document}

\begin{abstract}
Quantum computing promises to revolutionize our understanding of the limits of computation, and its implications in cryptography have long been evident. Today, cryptographers are actively devising post-quantum solutions to counter the threats posed by quantum-enabled adversaries. 
Meanwhile, quantum scientists are innovating quantum protocols to empower defenders. However, the broader impact of quantum computing and quantum machine learning (QML) on other cybersecurity domains still needs to be explored. 
In this work, we investigate the potential impact of QML on cybersecurity applications of traditional ML. First, we explore the potential advantages of quantum computing in machine learning problems specifically related to cybersecurity. Then, we describe a methodology to quantify the future impact of fault-tolerant QML algorithms on real-world problems. 
As a case study, we apply our approach to standard methods and datasets in network intrusion detection, one of the most studied applications of machine learning in cybersecurity. Our results provide insight into the conditions for obtaining a quantum advantage and the need for future quantum hardware and software advancements.
\end{abstract}
\begin{keyword}
Quantum computing \sep Quantum machine learning  \sep QML  \sep  Evaluation  \sep Framework  \sep Impact \sep  PCA  \sep  Principal Component Analysis  \sep  Network Intrusion Detection \sep  Network Security
\end{keyword}

\maketitle

\section{Introduction}
\label{sec:introduction} 

Quantum computing combines concepts from computer science, mathematics, and physics to provide a novel computational model. 
A \textit{quantum computer} is a programmable physical system that obeys the laws of quantum physics. Writing code for a quantum computer means specifying the evolution of a quantum mechanical system such that the system's final description encodes the output of the computation.
Expected to expand the way we process information radically, quantum computing provides new primitives unavailable in classical information processing. 

The impact of quantum computing extends across a wide range of domains. Quantum simulation, for example, allows researchers to model complex quantum systems that are intractable for classical computers, with significant implications for material science, chemistry, and fundamental physics. 
Additionally, quantum computing is driving the development of the quantum internet, which aims to utilize quantum entanglement for ultra-secure communication. 
The emerging field of quantum software engineering focuses on creating tools, languages, and frameworks to efficiently design and optimize quantum algorithms, ensuring practical and scalable implementations.

In cryptography, the influence of quantum computing has been particularly profound.
In 1994, Peter Shor published an algorithm for factoring integers and solving the discrete logarithm problem in polynomial time~\cite{shor1994algorithms}, mining the security of current asymmetric cryptography. Additional algorithmic primitives potentially threaten the security parameters of symmetric cryptography and hashing algorithms.
These developments profoundly impacted cryptography, leading to the dawn of post-quantum cryptography~\cite{bernstein2017post}, a research line striving to devise cryptographic primitives and protocols resilient to quantum-enabled attackers. 
Concurrently, quantum scientists are developing protocols to aid quantum-enabled defenders~\cite{pirandola2020advances}.
Cryptography, however, is not the sole domain expected to transform with the advent of practical quantum computing.

Another promising field is machine learning (ML), where quantum computers are expected to enable processing more extensive and complex datasets, potentially leading to more accurate and efficient algorithms. 
The interest in quantum machine learning (QML) can be traced back to 2009 when Harrow et al.~\cite{LLoyd} proposed a quantum algorithm to solve linear systems in time that depends only poly-logarithmically on the system's size and polynomially on other parameters. 
This result paved the way for new quantum linear algebra that proved very useful in ML~\cite{biamonte2017quantum}. 

Simultaneously, classical\footnote{In this manuscript, the word \quotes{classical} means non-quantum.} machine learning has been successfully applied to a wide range of cybersecurity tasks, from malware and binary analysis~\cite{ucci2019survey, rieck2011automatic} to detecting network intrusions~\cite{buczak2015survey} and financial frauds~\cite{Carminati2018}.
As ML algorithms become integral to both defenders' and attackers' arsenals for securing~\cite{arp2022and} and attacking~\cite{godefroid2017learn, yamaguchi2011vulnerability} computer systems, the potential impact of QML on cybersecurity becomes a natural question.

Recent literature has begun exploring the performances of near-term heuristic QML algorithms for computer security problems, such as denial of service or botnet detection.
However, the absence of theoretical guarantees for these near-term heuristics and the limitations of current hardware and simulators hinder researchers from predicting the scaling and performances of these algorithms on realistic datasets.
The research question of \textit{how QML can impact cybersecurity in the long run when fault-tolerant quantum computers are available} has been hardly explored in any previous literature to our knowledge.

In this work, we address this question by presenting a theoretical analysis of the potential advantages of quantum computing in ML. 
We build a simple framework to identify the conditions under which a QML algorithm for a fault-tolerant quantum computer outperforms a classical one. 
The goal of the framework is twofold: 
\begin{itemize}
    \item Study how the errors of QML algorithms --- which are a particular kind of randomized approximation algorithms --- affect the performance of quantum models with respect to their classical counterparts;
    \item Compare the running times of quantum and classical ML models to discover the settings where quantum algorithms provide an advantage. 
\end{itemize}

This framework enables cybersecurity experts and practitioners to assess whether quantum machine learning algorithms can effectively address specific practical problems. It provides a benchmarking tool for fair comparisons of future QML solutions in the cybersecurity domain, which will surely be proposed in the next few years.

As a demonstrative case study, we apply the methodology to PCA-based quantum machine learning models for network intrusion detection, a typical ML application in cybersecurity. 
In particular, we study how the algorithmic errors introduced during the QML training affect the performance of intrusion detection models on real datasets. 
To do so, we implement and test a numerical simulation of essential quantum subroutines -- amplitude estimation, phase estimation, and tomography of pure states. 
These subroutines are critical building blocks that enable simulations of many existing QML algorithms. 
In our case study, we evaluate the performances of quantum clustering (q-means~\cite{qmeans}) and quantum dimensionality reduction (QPCA~\cite{qadra}) for intrusion detection.

In summary, this paper contributes to elucidate the potential impact of quantum machine learning on cybersecurity by providing a theoretical framework for assessing the advantages of quantum algorithms over classical ones, as well as conducting a case study on PCA-based quantum algorithms for network intrusion detection, shedding light on their performance and feasibility in real-world scenarios.

The manuscript is structured as follows.
In Section~\ref{sec:qml algos}, we discuss the importance of machine learning in cybersecurity, the recent literature on near-term heuristic-based QML, and describe the opportunities for QML algorithms in the fault-tolerant regime.
Section~\ref{sec:quantum background} provides readers with the necessary quantum background to understand the proposed framework.
Section~\ref{sec:methodology} presents the evaluation framework.
Finally, in Section~\ref{sec:case study}, we present the case study on PCA-based quantum algorithms for network intrusion detection.


\section{Quantum machine learning algorithms for computer security}
\label{sec:qml algos}
In the rapidly evolving landscape of computer security, the intersection of quantum computing and machine learning has emerged as a frontier with transformative potential. As we witness the advancements in quantum technologies, researchers have embarked on developing quantum machine learning (QML) algorithms that could find cybersecurity applications. 
This section explores the connections between computer security and quantum machine learning, discussing the current landscape and the promising horizons within the fault-tolerant regime.

\subsection{Machine learning for cybersecurity}
Machine learning algorithms have found widespread success in cybersecurity applications~\cite{arp2022and,ucci2019survey,fraud_clust_1}. 

In the unsupervised learning domain, Principal Component Analysis (PCA) stands as an indispensable tool for preprocessing and analyzing large, complex datasets. 
Its applications span anomaly and intrusion detection, feature selection, and privacy protection. 
In anomaly detection~\cite{pca_traffic_anomaly_detection_1,pca_traffic_anomaly_detection_2}, PCA helps identify unusual patterns in network traffic, system logs, or user behavior, thereby making it easier to detect deviations that may signal cyberattacks or malicious activities. For Intrusion Detection Systems~\cite{pca_network_intrusion_1} (IDS) and Malware Detection, PCA aids in analyzing network traffic data and malware samples, respectively, and reduces the feature space to preserve the most pertinent information. 
Regarding feature selection in machine learning-based cybersecurity applications~\cite{pca_dim_red_network}, PCA selects relevant features, discarding less significant ones and enhancing the model's efficiency and accuracy.  
Additionally, the application of clustering algorithms (e.g., Hierarchical, k-means, and spectral clustering) has earned significant attention as an effective tool for identifying patterns and anomalies within large datasets of security-relevant phenomena like network intrusions~\cite{hier_clust_1, kmeans_clust_1, spectral_clust_1} and malware~\cite{ malw_clust_2}.

At the same time, supervised learning techniques such as Support Vector Machines (SVMs), linear regression, Neural Networks (NNs), and Convolutional Neural Networks (CNNs) are used for multiple tasks. 
SVMs find use in malware classification~\cite{svm_mal_1} and intrusion detection~\cite{svm_nn_intr_1}, as they can handle high-dimensional data and can segregate malicious activity from normal through hyperplane separation in feature space. 
Linear regression, though traditionally employed in predictive modeling, can also forecast the behavior of security metrics over time, thereby aiding proactive threat management. 
Neural Networks are used, for instance, in phishing detection~\cite{nn_phis} and vulnerability identification~\cite{nn_vuln}, thanks to their ability to learn complex patterns in data. Convolutional Neural Networks have proven to be particularly effective in malware analysis~\cite{cnn_malw_1, cnn_malw_2}. 

As machine learning establishes itself as a pivotal tool in cybersecurity, the potential applications of quantum machine learning (QML) become increasingly evident.
Accelerating computational processes through quantum computing has the potential to significantly enhance security measures, particularly in solving complex problems such as the analysis of massive-scale datasets, symbolic execution-based malware analysis, and automated vulnerability discovery through fuzzing.

\subsection{Current research trends}
In recent years, researchers have started exploring applying quantum machine learning to cybersecurity problems. 
Most ongoing applied research focuses on designing and testing QML algorithms for NISQ (Noisy Intermediate-scale quantum) devices -- small, non-error-corrected quantum computers.
This research line expects to ease the hardware requirements imposed by error-corrected architectures and start benefiting from the first quantum computers. 
The main idea is to program the quantum computer with parameterized gates, akin to classical neural networks, and optimize the circuit parameters using a classical computer. 
For a comprehensive review of the subject, we refer to~\cite{bharti2022noisy}. 

Along this research line, Kalinin and Krundyshev~\cite{kalinin2023security} evaluated the application of Quantum SVM and CNN to intrusion detection. 
Suryotrisongko and Musashi~\cite{suryotrisongko2022evaluating} investigated variational quantum algorithms for botnet detection based on domain generation algorithms (DGA). 
Payares and Martínez-Santos~\cite{payares2021quantum} discuss applications of QML using variational algorithms for SVM, Neural Networks, and an ensemble to detect denial of service (DOS) attacks. 
Masum et al.~\cite{masum2022quantum} use variational SVM and Neural Networks to investigate supply chain attacks.
Beaudoin et al.~\cite{beaudoin2022quantum} investigate using variational algorithms for quantum LSTM and NN to detect Trojan-infected circuits.

Despite the simplicity offered by these variational algorithms, obtaining provable guarantees about their performances remains challenging. 
Some recent results fuel the conjecture that it will be hard for parameterized quantum circuits to outperform significantly and consistently classical algorithms in many relevant areas~\cite{aharonov2023polynomial}. 
Moreover, the lack of theoretical guarantees makes quantum simulators and hardware crucial for evaluating their impact on practical applications, restricting the dimensions of the datasets on which it is possible to run experiments.
We believe that the proper test bench of these quantum algorithms will be when NISQ-scale quantum computers are available. 
The same difficulties stand for quantum annealing and machine learning approached through QUBO formulations~\cite{arthur2021qubo}.
For these reasons, we focus on QML algorithms with a provable speedup over their classical counterparts.

\subsection{Opportunities in the fault-tolerant regime}

Besides NISQ and QUBO algorithms, researchers have been investigating what fault-tolerant quantum computers equipped with a classically writable memory can do for machine learning.
Leveraging techniques from quantum linear algebra, researchers developed \quotes{quantized} versions of emblematic machine learning algorithms~\cite{biamonte2017quantum}.
They aim to reduce the running time requirements of the classical counterparts to process data in shorter times and analyze more data with the same computational power.
These algorithms come with theorems and analyses that bound their running times.
Examples of fault-tolerant machine learning algorithms with provable running times are the following.

\mypar{Unsupervised learning} There are several algorithms for unsupervised learning. 
Among the many, we cite quantum algorithms for hierarchical clustering~\cite{aimeur2013quantum},  spectral clustering~\cite{kerenidis2021quantumspectral}, nearest neighbor~\cite{wiebe2015quantum, lloyd2013quantum, bellante2023quantum}, k-means~\cite{qmeans}, Gaussian mixture models~\cite{kerenidis2020quantumgaussian}, quantum algorithms for PCA ed eigenvalue-based techniques~\cite{qadra}, and for learning sparse representations~\cite{qmp}.

\mypar{Supervised learning} The literature on supervised algorithms is equally proceeding. 
For instance, we have quantum algorithms for Support Vector Machines~\cite{rebentrost2014quantum}, linear regression~\cite{chakraborty2023quantum}, slow feature analysis~\cite{kerenidis2020classificationqsfa}, and many others. There are also some quantum algorithms for deep learning, like quantum neural networks, convolutional neural networks, and perceptron algorithms \cite{roget2022quantum, liao2021quadratic,kerenidis2019quantumQCNN}.
\smallskip

Most of these algorithms have the following important characteristic: once provided quantum access to the data, the complexity of the quantum algorithms depends only polylogarithmically on the number of data points. 
This starkly contrasts with classical algorithms, where the dependence is at least linear. 
The theoretical guarantees that come with analyzing these algorithms make them suitable for evaluating the impact of quantum computing in the future, allowing researchers to investigate their use on realistic datasets without the need for quantum simulators or hardware (e.g., see experiments in \cite{qadra, kerenidis2020classificationqsfa, dalzell2023end}). 
For this reason, we will focus on evaluating the impact of quantum machine learning algorithms in the fault-tolerant regime.
In the reminder, we use QML to refer to quantum machine learning algorithms with provable guarantees in fault-tolerant settings.

\subsection{Motivation and relevance to cybersecurity}
The cybersecurity field faces growing challenges from increasingly sophisticated threats, requiring continual advancements in detection and mitigation techniques. While traditional machine learning algorithms have demonstrated considerable success in areas such as intrusion detection, anomaly detection, and malware analysis, these methods are increasingly strained by the scale and complexity of modern security problems. Quantum machine learning offers a potential pathway to overcoming these limitations by enabling computational speedups and enhanced algorithms that could significantly improve cybersecurity defenses.

The primary motivation for this work is to explore the future implications of fault-tolerant quantum machine learning algorithms in cybersecurity applications. 
To date, much of the research in quantum computing for cybersecurity has centered around NISQ devices, which, while promising, are constrained by hardware limitations and lack theoretical guarantees.
These limitations hinder their scalability and applicability to large, real-world datasets, such as those commonly used in cybersecurity.
Our approach is different: we focus on QML algorithms designed for fault-tolerant quantum computers, which hold the potential to overcome these constraints and deliver provable performance guarantees.
In this paper, we define a quantum algorithm as advantageous if it runs faster than the classical version while still delivering comparable results.
However, the exact conditions under which quantum algorithms outperform their classical counterparts remain uncertain, particularly in practical cybersecurity applications. This uncertainty presents both a challenge and an opportunity for research.

Our work aims to address this gap by providing an evaluation framework that allows cybersecurity experts to systematically assess the potential advantages of QML algorithms in a fault-tolerant setting.
We also offer a case study on PCA-based anomaly detection for network intrusion that illustrates how such a framework can be applied to standard machine learning tasks, offering insights into when and how quantum algorithms may achieve a meaningful advantage.
While the case study serves to showcase the framework, the broader contribution lies in the methodology, which can be adapted to various cybersecurity domains and tasks.
By focusing on fault-tolerant quantum computing and algorithms with provable guarantees, this paper establishes a foundation for future research that will be relevant as quantum hardware advances. The work not only highlights the potential impact of QML on cybersecurity but also equips practitioners with practical tools to evaluate this impact, ensuring that the cybersecurity community is prepared to harness the capabilities of quantum computing when the technology matures.
In summary, the motivation behind this research is twofold: first, to investigate the theoretical underpinnings and performance benefits of fault-tolerant QML algorithms in cybersecurity; and second, to provide a robust framework for practitioners to assess the practical feasibility and scalability of these algorithms as quantum computing technology continues to evolve.


\section{Quantum computing background}
\label{sec:quantum background}

This section provides an overview of the key quantum computing concepts essential for evaluating fault-tolerant quantum machine learning (QML) algorithms.
Readers seeking a brief introduction to quantum computing are directed to \ref{apx:intro quantum}, while a comprehensive one is available in~\cite{nielsen2010quantum}.

\mypar{Qubits and registers} 
A qubit, or quantum bit, is the fundamental unit of quantum information. 
Differently from classical bits, qubits can exist in a superposition of $\ket{0}$ and $\ket{1}$ states and collapse to one of these states when measured.
The state of a single qubit is described by a complex vector with two entries, from which it is possible to compute the probability of collapsing in one state or another. 
The state of multiple qubits can be described by a complex vector of size exponential in the number of qubits, and qubits can be correlated.

\mypar{Quantum algorithm} A quantum algorithm is a computational procedure that consists of two key processes: evolving and measuring a specific initial quantum state multiple times. 
The evolution process involves applying a series of quantum operations or gates to the initial quantum state. 
These operations change the state of qubits in a controlled manner. 
Instead, measurement collapses the quantum state from its superposition of outcomes to a definite state, yielding the outcome of the computation. 
The process of reconstructing a quantum state requires statistics over several measurements and is called \emph{tomography}. 
To perform tomography, one must re-create the quantum state by repeating the algorithms from scratch multiple times (e.g., Theorem~\ref{thm:tomography}).

\mypar{Notation} We adopt Dirac's notation, denoting a complex vector $\boldsymbol{x} \in \C^n$ as $\ket{x}$, for some $n \in \N$, and its transposed complex conjugate $x^\dagger$ as $\bra{x}$. 
Tensor products $\ket{x} \otimes \ket{y}$ are abbreviated as $\ket{x} \ket{y}$.
For a matrix $\boldsymbol{X} \in \R^{n\times d}$, $\norm{\boldsymbol{X}}$ represents the spectral norm (the greatest singular value), $\kappa(\boldsymbol{X})$ denotes the condition number (the ratio between the greatest and the smallest singular values), and $\mu(\boldsymbol{X})$ is a parameter bounded by the Frobenius norm (i.e., $\mu(\boldsymbol{X}) \leq \|\boldsymbol{X}\|_F = \sqrt{\sum_{ij}X_{ij}^2}$, see Def.~\ref{def:mu} in \ref{app:math}).
Big-$O$ notation is used for algorithmic complexity, with $\Ot$ omitting polylogarithmic terms.

\subsection{Mapping quantum software on hardware}
\label{sec:mapping}

At the time of writing, numerous research institutions and companies invest considerable resources in developing quantum computers, with some prototypes already available in the cloud. 
Current quantum computers have qubits in the order of hundreds and can execute gates in the order of thousands. 
These prototypes have neither enough qubits nor enough quantum coherence to run any useful computation with a provable advantage over classical computers for problems of real interest. 
Nevertheless, in the past years, some hardware architectures have successfully solved computational problems of purely theoretical interest (i.e., with no practical application) faster than any classical computer. 

When a quantum hardware platform can execute a task that is computationally infeasible for classical computers, it is said to have achieved \emph{quantum supremacy} or \emph{quantum advantage}~\cite{madsen2022quantum,preskill2012quantum}. 
It is reasonable to expect that in the next decades, quantum computers will be mature enough to execute software for more relevant problems. 
Interested readers are referred to~\cite{saffman2010quantum, bruzewicz2019trapped, kjaergaard2020superconducting} for an overview of scalable quantum computer architectures. 
While quantum architectures are expected to become faster and more reliable over time, they are not projected to surpass classical architectures in clock time~\cite{babbush2021focus}. 
Consequently, considerable speedups must come from algorithmic improvements in the number of operations executed.

In the reminder, we describe the steps to consider when compiling quantum software into quantum hardware. 
All these steps add some overhead that needs to be considered when estimating the wall-clock time of a quantum algorithm.

\mypar{Programming} 
From a programmatic standpoint, conceptualizing quantum computers is akin to thinking of an FPGA, where the circuit is described for mapping and execution on the device.
To facilitate the development of quantum software, quantum computers can be programmed in high-level languages. 
Although these languages are less advanced than the ones we can use for classical hardware, they facilitate specifying abstract gates and circuits (e.g., arbitrary qubit rotations, adders, multipliers) and common subroutines such as quantum Fourier transform and amplitude amplification and estimation. 
Often, the code generated by these high-level languages is not optimized for the task or the target hardware. 

\mypar{Compilation} 
Quantum hardware platforms typically execute only a basic set of universal quantum gates, i.e., gates that can be combined to build any other circuit.
These gate sets often vary from architecture to architecture. 
During compilation, high-level instructions are decomposed into sequences of these gates.
The Solovay-Kitaev theorem enables efficient transpilation of quantum software between architectures, though this process can introduce some overhead in the total number of gates~\cite{maronese2022quantum,bouland2021efficient}. 
Optimization techniques based on heuristics are available~\cite{duncan2020graph} and can help reduce the circuit size and depth. 
Besides dealing with circuit decomposition and optimization, compilation will also need to take into account error correction and connectivity constraints.

\mypar{Error correction} 
The primary challenge in achieving large-scale quantum computers is constructing noise-resilient qubits.
Quantum Error Correction Codes (QECC) are a set of methods that help protect quantum information from noise. 
They usually embed one logical qubit in a bigger Hilbert space spanned by many physical qubits and work by detecting errors and applying corrective gates at runtime. 
QECC introduces overhead in terms of qubit quantity, number of operations, and classical interaction needed to decode the errors and control the quantum processing unit accordingly. 
With current technology, this overhead is such that asymptotic quadratic speedups are believed insufficient for practical scenarios~\cite{babbush2021focus}.

\mypar{Connectivity} 
Another factor influencing compilation and effective quantum algorithm running time is hardware connectivity. 
In fact, some architectures limit connectivity to interactions between physically proximate qubits. 
To overcome this limitation, qubits must be swapped along the circuit using intermediate qubits to build long-range interactions.
This introduces an overhead in the number of operations. 
During compilation, logical qubits are assigned to physical qubits to minimize long-range interactions, as connectivity constraints may limit circuit parallelization.

\subsection{Complexity of quantum algorithms}
\label{sec:complexity}

There are different complexity measures for quantum algorithms. 
As in the study of classical algorithm complexity, at a theoretical level, we are interested in the \stress{asymptotic} scaling in the problem parameters.

The \emph{query complexity} of an algorithm is the number of calls made to an \emph{oracle} providing access to the problem's input.
This measure -- which is also standard in classical computer science -- conceals the oracle's implementation cost and the processing cost between any two oracle calls. 
One of the reasons for its adoption in quantum (and classical) computer science is the use of techniques like the polynomial and adversarial methods to prove lower bounds in this model~\cite{ambainis2000quantum,beals2001quantum}.

\begin{figure}[t]
\centering
\begin{tcolorbox}[
boxrule=0.3mm,
colback=black!5,
width=(\linewidth),before=\hfill,after=\hfill,]
\mypar{Example 1 - Query advantage $\Rightarrow$ gate advantage} \\
Consider a quantum application targeting a preimage attack scenario, where an oracle provides access to a hash function (e.g., $\mathtt{SHA256}$), comparing the output with a target hash image.
The brute-force query complexity, evaluated in terms of hash function calls, is on the order of $O(2^m)$ for an $m$-bit hash function. 
Using Grover's search algorithm~\cite{grover1997quantum} with a quantum computer,  a quantum computer reduces this complexity to $O(2^{m/2})$.
Notably, this result only suggests a reduction in the number of hash function evaluations.
However, in this case, an advantage in query complexity translates into an advantage in gate complexity, as the costs of the quantum and classical oracles are comparable and vastly dominate the cost of intermediate operations between oracle calls.
To express the quantum cost in terms of time complexity, one must delve into the implementation details of the oracle (i.e., a quantum circuit for $\mathtt{SHA256}$) and consider the overhead for error correction. 
An analysis of the physical resources (the number of qubits and the actual number of gates) and wall-clock time estimation, with surface error correction codes on planar connectivity, can be found in~\cite{amy2016estimating, aggarwal2018quantum}.
\end{tcolorbox}
\end{figure}

The \stress{gate complexity} denotes the total number of one and two-qubit gates executed by the quantum circuit implementing the algorithm. 
This choice stems from the fact that any multi-qubit gate can be decomposed into a sequence of one- and two-qubit that form a universal gate set and that the asymptotic behavior of the gate complexity remains unaffected by the choice of the specific gate set.
This complexity measure is the quantum equivalent of a classical Boolean circuit's \emph{gate complexity} or the \stress{circuit-size complexity}.
Given the query complexity and a gate decomposition of both the oracle and the intermediate operations between the oracle calls, one can determine the gate complexity.

The \stress{depth} of a quantum algorithm mirrors the classical \emph{circuit-depth complexity} of a Boolean circuit. 
It is the length of the longest sequence of gates from the input to the output of a quantum circuit. 
Typically measured before circuit compilation on specific hardware architecture, it does not consider Quantum Error Correction (QECC) or connectivity issues. 
Depth complexity offers insight into the degree of parallelization achievable within a quantum algorithm.

The \emph{time complexity} gauges the wall-clock time required to execute the algorithm on specific hardware. 
Unlike the previously mentioned asymptotic complexity measures, which abstract away implementation details, this metric reflects the actual execution time of a quantum algorithm on the chosen hardware platform, which may impose constraints on parallelism, available error-correcting codes, and compilation techniques. 
An estimate of the time complexity can be derived from the query complexity, along with the details on the implementation of the oracles and the details of the hardware architecture (such as compilation, error correction, connectivity, and parallelization). 
The process of estimating the time complexity of a quantum algorithm by expanding the oracle, optimizing the circuit, and considering all the hardware and technological constraints is called \emph{resource estimation}.

We clarify the previous definitions with two simple examples. 
In Example~1, we observe that query complexity serves as a reliable measure to assess the efficiency of a quantum algorithm, as the advantages readily extend to gate and time complexity.
Contrastingly, Example~2 emphasizes the need for caution when relying solely on query complexity. 
The quantum algorithm for the hidden subgroup problem exhibits a stark difference between query and gate complexity, challenging the assumption that reducing query complexity guarantees efficiency gains in other aspects.

Researchers commonly express the complexity of fault-tolerant quantum machine learning algorithms in terms of queries to input oracles.
Analogous to Example~1, in QML, the negligible cost of intermediate operations between oracle calls often positions memory accesses as the bottleneck. 
This aligns with the understanding that, in quantum machine learning, query complexity remains a pertinent metric, offering valuable insights into the algorithm's efficiency.

\begin{figure}[t]
\begin{tcolorbox}[
boxrule=0.3mm,
colback=black!5,
width=(\linewidth),before=\hfill,after=\hfill,]
\mypar{Example 2 - Query advantage $\centernot{\Rightarrow}$ gate advantage} \\
An illustrative example highlighting a significant disparity between query complexity and gate complexity is the quantum algorithm for the hidden subgroup problem~\cite{ettinger2004quantum}.
In this case, the algorithm requires only a polynomial number of queries to the function oracle, seemingly indicating efficiency.
However, the number of gates between successive calls results in an exponential gate complexity.
For this algorithm, query complexity is not a good proxy for time complexity.
\end{tcolorbox}
\end{figure}

\subsection{Classical data and quantum computers}
\label{sec:classical data quantum computers}
To process classical data in a quantum computer, we must consider input and output operations, which we survey in this section.

\mypar{Input}
Along with a quantum computer, it is common to assume the availability of a classically-writable quantumly-readable quantum random access memory (QRAM).
Here, a QRAM is a memory that can be written with a classical device and responds to queries in superposition from a quantum device. 
A memory $[m_0, \dots m_i, \dots m_{M-1}]$ is an indexed array of size $M$, with entries encoded in $p$ bits.
A quantum memory is a unitary of the kind 
\begin{equation}\label{eq:queryqram}
U_{QRAM}: \ket{i}\ket{0}\mapsto \ket{i}\ket{m_i},
\end{equation}
which maps the $i^\mathrm{th}$ address to the memory entry $m_i$.

Some algorithms may require the binary encoding of $m_i$ in a register of qubits, as produced by the unitary in Eq.~\ref{eq:queryqram}.
Others may need to encode the entries in the amplitudes of a quantum state $\ket{m} = \frac{1}{\norm{m}}\sum_{i=0}^{M-1} m_i \ket{i}$.
Others again require access to a unitary $U$ (a circuit) such that $\norm{A - \alpha (\langle 0|^{\otimes q}\otimes I ) U (|0\rangle^{\otimes q}\otimes I)} \leq \epsilon$, where $A$ is a matrix representation of a portion of the memory (i.e., $(\alpha, q, \epsilon)$-\emph{block-encoding} access to a matrix~\cite{low2019hamiltonian}).
Efficient access to all these data representations can be achieved with a QRAM oracle, as described in Eq.~\ref{eq:queryqram}.

While a basic multiplexer circuit can implement the mapping in Eq.~\ref{eq:queryqram}, it would have a linear depth in memory size  $O(Mp)$.
There exist more efficient circuit implementations for creating access to sparse data structures, such as sparse vectors or matrices~\cite{camps2024explicit,di2020fault}.
Fortunately, better general architectures for implementing Eq.~\ref{eq:queryqram} exist.
For instance, the bucket-brigade architecture maintains a total circuit size of $O(Mp)$ gates but has a depth only logarithmic in the memory size~\cite{giovannetti2008architectures} and is quite resilient to generic errors~\cite{hann2021resilience}. 
Although the first log-depth QRAM architecture was presented in 2008, quantum random access memories only recently started to become an active topic in the research community. 
For instance, different research proposals exist for implementing this architecture on a different kind of quantum hardware~\cite{hann2019hardware} and in data centers~\cite{liu2022quantum}.

We have \emph{efficient access to a quantum memory} if the mapping in Eq.\ref{eq:queryqram} can be performed in time $O(\mathrm{poly}(\log (M), p))$.

\mypar{Output}
To retrieve data out of a quantum computer, the possibilities are more narrow. 
In QML, the output is often encoded in the amplitudes of a quantum state that the algorithm produces.
Subroutines like amplitude estimation (Theorem~\ref{thm:amplitude estimation}) can be used for a single-number output.
In cases where the output is a vector, one can use quantum tomography of pure states (Theorem~\ref{thm:tomography}). 
Quantum pure-state tomography enables the reconstruction of a quantum state by sampling it. 
Given a quantum state $\ket{\boldsymbol{x}}$, the tomography algorithm outputs an estimate $\overline{\boldsymbol{x}}$ such that $\lVert \boldsymbol{x}-\overline{\boldsymbol{x}} \rVert_2\leq\delta$ by performing $N=\widetilde{O}\left(\frac{d}{\delta^2}\right)$\footnote{Note that tomography subroutines have been recently improved in general settings, offering a further quadratic speedup in the precision $\delta$~\cite{van2023quantum}.} measurements~\cite{tomography}. 


\section{Evaluating the quantum advantage}
\label{sec:methodology}

This section introduces a framework designed to assess the quantum advantage in machine learning applications. We focus on cybersecurity problems. 
However, the methodology is general and can be adapted for evaluating the impact of QML algorithms in various fields. 
First, we analyze fault-tolerant quantum machine learning algorithms, emphasizing the challenges posed by the data loading problem and the complexities in estimating their running times on real-world datasets. Subsequently, we detail the methodology framework, discussing its merits and limitations.

\subsection{Fault-tolerant quantum machine learning}
We discuss the caveats arising from the cost of loading classical data into a quantum device and the challenges associated with evaluating the complexity and impact of a quantum machine learning algorithm.

\subsubsection{The cost of data loading}
Quantum machine learning algorithms can process both native quantum data, originating from controllable quantum algorithms and processes, and classical data, derived from conventional sensors and machinery. 
When assessing the cost of processing classical data, one must also consider the expenses associated with compiling the unitaries that provide input to the data.

In applying QML to contemporary cybersecurity challenges, like network intrusion detection or malware analysis, the criticality of the classical data loading cost onto quantum devices becomes evident. 
In some cases, refraining from considering this cost by assuming that the data is pre-stored in some quantum-accessible memory may lead to inaccurate assessments of the quantum algorithms' efficiency, as emphasized by Aaronson~\cite{aaronson2015read}.

QML algorithms capitalize on the ability to store a size $d$ vector using only $\lceil \log d \rceil$ qubits and an $N\times d$ matrix using $\lceil \log N \rceil + \lceil \log d \rceil$ qubits.
Amplitude encodings or block-encodings facilitate this storage. 
With efficient quantum memory access (Eq.~\ref{eq:queryqram}) and some dataset preprocessing, these encodings can be generated in polylogarithmic time in the matrix dimensions. Detailed methods for leveraging QRAM for this purpose are outlined in \ref{app:math}.

It is important to highlight that these encodings necessitate preprocessing the input data on a classical device. 
This step consists of preparing a specific data structure to store in the quantum memory rather than the raw matrices or vectors. 
The preprocessing time for a $N \times d$ matrix is  $\widetilde{O}(Nd)$ and is executed only once during loading. 
This process is highly parallelizable, efficient to update ($\widetilde{O}(1)$ per matrix entry), and is an essential factor in evaluating the quantum algorithm's speed compared to classical methods. 

Considering this preprocessing input step, the expected exponential speedup of the quantum procedure is often lost, as one needs to pay a linear time data loading~\cite{aaronson2015read}.
As an example, this data loading cost implies that quantum algorithms may not significantly expedite the inference phase of classical machine learning models whose prediction time is linear in the number of features. 
In such cases, the time required to input the test data point into a quantum memory would be comparable to the prediction time itself.
Nevertheless, in situations where a model's training or inference phase incurs a polynomial time cost, quantum algorithms could still offer a comparative polynomial advantage over classical alternatives. 

\subsubsection{Complexity evaluation}
\label{sec:complexity evaluation}

The complexity of fault-tolerant quantum machine learning algorithms is often expressed in terms of \emph{queries} to a quantum memory or in the total number of \emph{gates} to execute considering efficient access to a quantum memory (i.e., polylogarithmic, similar to the memory access cost in the classical RAM model).
Similar to Example 1 -- unlike Example 2 -- in QML algorithms, these two costs are equivalent up to polylogarithmic factors. 
This equivalence makes the QRAM query complexity a robust metric for quantifying the number of simple operations required. 
We leverage this fact in our evaluation methodology, enabling decisions based on query complexity before executing detailed resource estimations.
The QML algorithms considered here are \emph{randomized approximation algorithms}.
Their running times depend on some probability of failure $\gamma$ and approximation error $\epsilon$. 
It is possible to trade off these parameters to expedite the algorithm's execution at the expense of performance and reliability.
A cybersecurity expert might want to tailor the amount of tolerable error and failure probability to the problem they are trying to solve, striking the best trade-off between time efficiency and accuracy/reliability.

\begin{figure}[th!]
\centering
\begin{tcolorbox}[
boxrule=0.3mm,
colback=black!5,
width=(\linewidth),before=\hfill,after=\hfill,]
\mypar{Example 3 - Approximation error vs Running time} \\
Consider an anomaly detection system based on Euclidean distances, where a test vector $\vec{x}_t$ is flagged as an anomaly if it falls within a radius $r$ of vector $\vec a$, i.e., $\|\vec{x}_t - \vec a\|_2 \leq r$.  
For vectors of $d$ features, an exact and deterministic classical classifier would require $O(d)$ operations.
On the other hand, a quantum  classifier that uses Theorem~\ref{thm:innerproductestimation} can evaluate the distance $D$ with an error $\epsilon$ (i.e., estimates $\overline{D} \in [D-\epsilon, D+\epsilon]$ with high probability) with query complexity and extra number of gates scaling proportionately with the inverse error ($\Ot(\frac{1}{\epsilon})$). 
Determining a suitable value for $\epsilon$ involves striking a balance between speed and classification accuracy. 
If the error is too big, we risk misclassifying anomalous vectors, e.g., $D\leq r$ but $D+\epsilon > r$. 
On the other hand, if the required error is too small, we end up with a quantum algorithm slower than its classical counterpart.
In conclusion, the quantum algorithm seems advantageous whenever the problem tolerates an error that is larger than the inverse number of features $(\epsilon \in \Omega(1/d))$.
Conversely, the classical algorithm seems advantageous whenever the problem requires an error smaller than the inverse number of features $(\epsilon \in \Ot(1/d))$.
Overall, the advantage is not self-evident nor easy to evaluate, as it depends on the problem and its data.
\end{tcolorbox}
\end{figure}

\begin{figure}[th!]
\centering
\begin{tcolorbox}[
boxrule=0.3mm,
colback=black!5,
width=(\linewidth),before=\hfill,after=\hfill,]
\mypar{Example 4 - Dataset parameters}\\ 
Consider a quantum and a classical algorithm for fitting a PCA machine learning model. 
The model consists of the first $k$ right singular vectors, corresponding to the largest $k$ singular values of a matrix $\boldsymbol{X} \in \mathbb{R}^{n \times d}$. 
The running times of the classical and the quantum algorithms are:
\begin{equation*}
\label{eq:runtime pca}
    \resizebox{\linewidth}{!}{    
    $\widetilde{O}_{\rm{c}}\left(\frac{kdn}{\sqrt{\epsilon_L}} \log(\frac{1}{\gamma})\right) ~ 
    \widetilde{O}_{\rm{q}}\left(\frac{kd}{\delta^2}
    \frac{\norm{\boldsymbol{X}}}{\theta}\frac{1}{\sqrt{p}}\frac{\mu(\boldsymbol{X})}{\epsilon_Q}
    \log(\frac{1}{\gamma})\right)$
    }
\end{equation*}
Here, $\epsilon_L$ represents the error related to the relative spectral gap between eigenvalues, $\theta, \epsilon_Q$ and $\delta$ are the quantum algorithm approximation errors, $p$ is the amount of retained variance, $\mu(X)$ as per Def.~\ref{def:mu} and $\gamma$ is the failure probability. 
Note that the dependence on $n$ is encapsulated in the $\widetilde{O}$ notation, as the quantum algorithm depends only polylogarithmically on it. 
One needs to estimate the parameters governing the query complexity of the QML algorithm on real datasets and use these estimates to compare the performances of the two algorithms.
\end{tcolorbox}
\end{figure}

While theorems provide asymptotic query complexity for QML algorithms, comparing this complexity to classical algorithms is not straightforward. 
Indeed, quantum and classical running times often depend on different dataset parameters. 
For instance, a classical algorithm for machine learning might depend solely on the dataset's size. 
In contrast, the query complexity of its quantum counterpart might depend on the effective condition number $\kappa(\boldsymbol{X})$, some approximation error $\epsilon$, a failure probability  $\delta$, one choice of $\mu(\boldsymbol{X})$ (Def.~\ref{def:mu}), and other dataset-dependent parameters.
These dataset-dependent parameters are critical in evaluating regimes in which the quantum algorithm can exhibit advantages over its classical counterparts (see Examples 3 and 4).

\mypar{Failure probability} 
QML algorithms fail with a probability smaller than $\gamma > 0$, similar to classical randomized algorithms. This probability can often be minimized by incurring in a multiplicative cost in the running time of $O(\log (1/\gamma))$, which is negligible in practice.
Once the cybersecurity expert fixes the tolerable failure probability based on the application, this factor accounts for a constant in the total running time.

\mypar{Approximation error} The output of QML algorithms approximates the output of corresponding classical subroutines. 
For a vector output $\overline{s}$, we consider an $\ell_2$ or $\ell_\infty$ approximation error over the classical output $s$, i.e., a vector $\overline{s}$ such that $\lVert s-\overline{s}\rVert_2\leq\epsilon$ or $\lVert s-\overline{s}\rVert_\infty\leq\epsilon$. 
In the case of a scalar $a$, we consider the absolute or relative error between the classical and the quantum output, i.e., $\lVert a- \overline{a} \rVert \leq \epsilon$ or $\lVert a- \overline{a} \rVert \leq \epsilon a$. 

\mypar{Other dataset-dependent parameters} Several parameters impact QML algorithms' performance, such as the maximum norm of vectors in the dataset (see Theorem~\ref{thm:innerproductestimation}, or $\eta$ in Theorem~\ref{thm:q-means}, or $s_q$ in Def.~\ref{def:mu}), the sparsity of the matrix, or a threshold $\theta \in (0,\norm{\boldsymbol{X}}]$ for picking the chosen number of principal components (Theorem~\ref{TheoMio:top-k_sv_extraction}).
Another critical quantity is the condition number of the dataset matrix.
In many real-world applications, matrices are not full-rank, meaning some small singular values are zero or nearly zero, potentially resulting in an unbounded condition number. 
Discarding singular values below a threshold reduces the effective condition number, potentially enhancing numerical stability and algorithm performance.
The threshold depends on the dataset, the algorithm, and the application, and its optimization can benefit the numerical stability and the performances of the (Q)ML algorithm, as it can help regularize the model.

\subsection{The Evaluation Framework}
\label{methodology}

\begin{figure}[t]
    \centering
    \includegraphics[width=0.8\columnwidth]{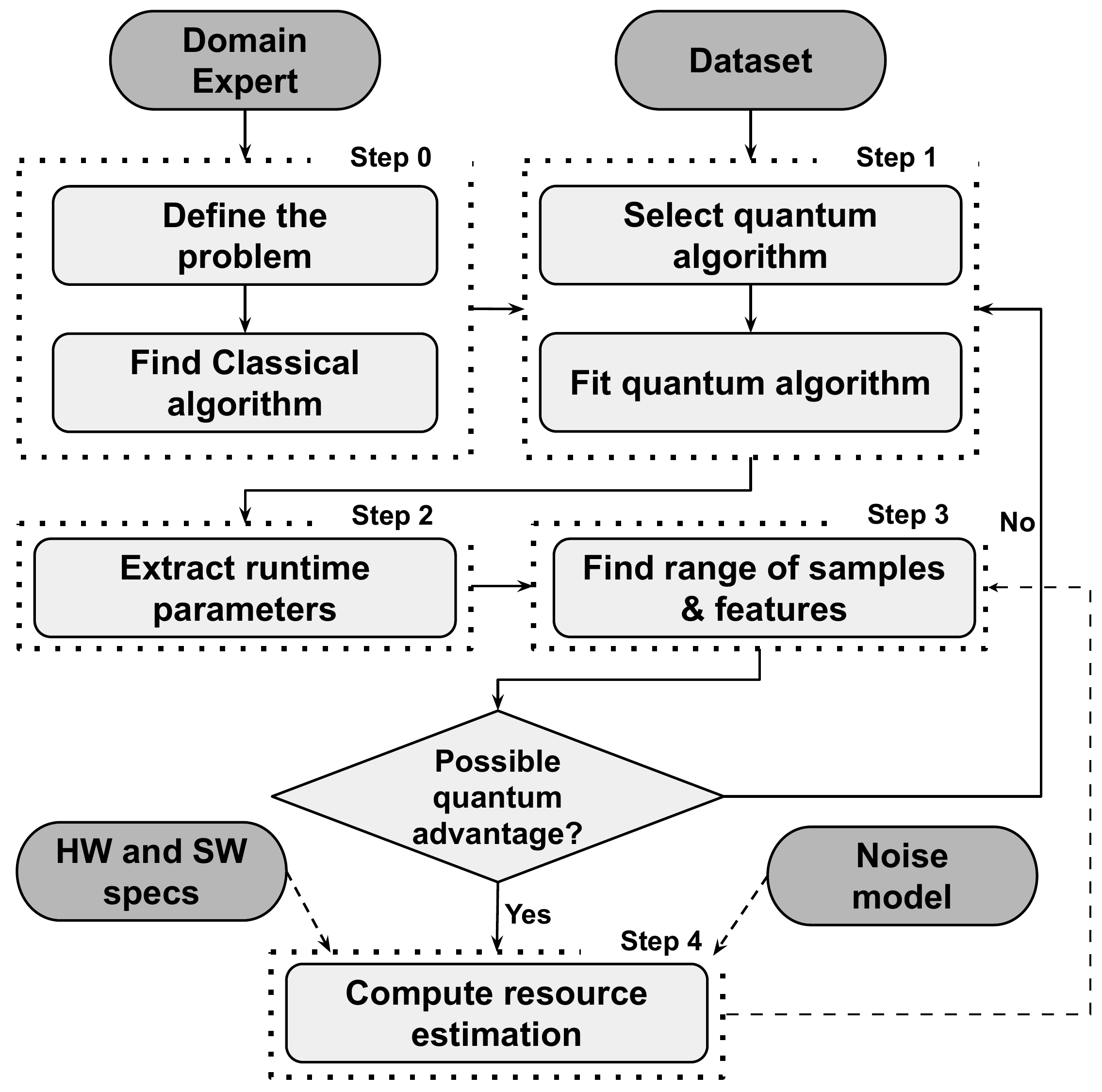}
    \caption{Framework for evaluating speedups with quantum algorithms. 
    }
    \label{fig:methodology flowchart}
\end{figure}

Building on the considerations of the previous sections, we describe a framework for evaluating the impact of quantum machine learning algorithms on cybersecurity problems.
The framework is summarized in Figure~\ref{fig:methodology flowchart}.

\myparb{Step 0} \emph{Formally define the problem and identify the best classical algorithm.} Start by defining a machine learning problem.
Collaborate with experts to formalize the problem, select the best classical algorithm, and choose a representative dataset for assessing speedup. 
The best classical algorithm might be the one with the best asymptotic complexity or the best performance in practice. 

\myparb{Step 1} \emph{Select a quantum machine learning model. Train the model and maximize the amount of approximation errors. 
Stop before the performance of the QML algorithm gets worse than tolerable}.
\textcircled{a} Select a candidate quantum machine learning algorithm to solve the problem. The quantum algorithm may not be present in literature or may need to be tailored to the problem, requiring the help of a quantum algorithm researcher.
\textcircled{b} Model the quantum algorithm with a classical algorithm that simulates the quantum procedure by artificially adding errors in the relevant steps, following the error types expected from the quantum algorithm's theoretical structure. 
Design the error to realistically/pessimistically model the approximation error expected in the actual quantum computation.
Fit the quantum model and find the best set of parameters (e.g., approximation errors, effective condition number) that enables a satisfying performance, minimizing the running time.
The parameters can be found manually or using algorithms for hyper-parameter tuning. 
Usually, a satisfying performance matches (or improves) the classical counterparts. 

\myparb{Step 2} \emph{Measure the dataset parameters that influence the running time of the quantum algorithm.} Measure the other dataset parameters that influence the quantum algorithm's running time and cannot be traded for time efficiency.
Examples can be the Frobenius norm of the dataset matrix, the maximum norm of the data points, the condition number, or other parameters specified in the theorem of the quantum algorithm and described in Section~\ref{sec:complexity evaluation}.

\myparb{Step 3} \emph{Find the combinations of the number of data points and features that enable quantum advantage.} 
Use the parameters found in the previous steps to quantify the query complexity of the quantum algorithm and the classical complexity as the number of data points and features increase. 
Estimate the dataset size (number of samples and features) at which the QML algorithm offers a significant advantage in query complexity compared to the classical approach.

\myparb{Decision} If a practical dataset size is identified where the quantum algorithm exhibits an advantage in query complexity, proceed to step 4. Otherwise, return to step 1 and consider using a classical algorithm if no suitable quantum algorithms are found.

\myparb{Step 4} \emph{In-depth resource estimation}. 
Select a hardware architecture with a given connectivity and gate set, a noise model of the qubits and gates, and an error correction code.
Estimate the number of qubits and the resources needed for a better wall-clock time than the classical algorithm. 
In this context, the resource overhead associated with achieving fault tolerance — namely, the increased number of physical qubits, execution time, and overall cost — must be critically evaluated. If the required execution time per quantum gate is deemed unrealistic due to these overheads, reconsider the dataset dimensions identified in step 3.

\smallskip

This methodology can be used to \emph{identify} promising QML applications and \emph{rule out} the tasks for which a quantum advantage cannot be reasonably expected.
It can be used to evaluate speedups both in the \emph{training} and \emph{inference} phases of QML algorithms. 
Under the assumption that the quantum clock-time will not become faster than the classical one~\cite{babbush2021focus} and that the theoretical analyses of the algorithms are tight, the Decision step after step 3 suffices to understand which tasks might benefit from quantum advantage in the future.
However, this decision process must consider the potential resource overhead introduced by fault-tolerant requirements.
Once it is established that a quantum advantage might be plausible for a task, an \emph{in-depth resource estimation} will provide further insights into the algorithm feasibility within the current state of the technology and into the necessary hardware improvements that could enable the advantage in the future.
While we expect that improving quantum hardware will enable the practicality of many quantum advantages, the reader should also be mindful that classical architectures and computers are expected to improve, albeit at a slower pace.

\subsubsection{Advantages and limitations}
\mypar{The Decision step} This conceptual framework's most significant advantage and limitation lie in the \emph{Decision} step.

On the positive side, ruling out the advantage of a QML algorithm based on its QRAM query complexity might spare the researchers the burden of performing an in-depth resource estimation, enabling cybersecurity experts and practitioners to expedite their evaluation of QML algorithms. 
Indeed, the parameter estimation steps (1-2) can be performed without the need for quantum simulators, enabling theoretical studies on large and realistic datasets. 
Rather than simulating the algorithm on a quantum simulator, practitioners will need to understand how to introduce artificial errors in the classical versions of the quantum algorithms, following the theoretical analysis of the quantum algorithm.
Performing in-depth resource estimations requires deep knowledge of quantum technologies, and the scientific effort needed to make a proper one makes for a scientific contribution on its own. 
A preliminary analysis of the advantage in query complexity can be enough to understand the potential of a QML algorithm and spare the need for detailed resource estimation.

On the negative side, to successfully rule out an algorithm at the decision step, we need the following assumptions:
\begin{itemize}
\item The quantum computer's clock will not significantly outpace a classical one.
\item The asymptotic query complexity is tight (and not loose because of proof artifacts).
\item The models for the simulated errors align realistically with actual errors. 
\end{itemize}

Any deviation from these assumptions risks discarding a valuable candidate for quantum advantage.
Consequently, the Decision step makes the evaluation framework a cautious approach, potentially missing advantageous QML candidates while avoiding misclassifying a suboptimal quantum algorithm as advantageous.

\mypar{Metrics for advantage} The criteria for selecting a quantum algorithm in \emph{Step~1} could benefit from further elaboration.

\emph{In this work, a quantum algorithm is considered advantageous if it demonstrates superior speed compared to its classical counterpart while retaining comparable performance.}
However, the selection criteria might be based on more complex security properties of QML algorithms, such as their robustness and resilience to adversarial attacks. 
As of today, the robustness properties of many QML algorithms still need to be explored and require further research.
This broader perspective could enhance the evaluation's depth, accounting for security beyond raw computational efficiency.


\section{Case study: PCA-based network IDSs} \label{sec:case study}

In this section, we demonstrate the evaluation framework on PCA-based QML methods and realistic datasets in one of the most studied applications of ML in cybersecurity: network intrusion detection.
Specifically, we investigate potential quantum speedups in the training phase of three PCA-based anomaly detection algorithms: the Principal Component Classifier (PCC) by Shyu et al.~\cite{PcaMajorMinor}, the Reconstruction Loss method used in Verkerken et al.~\cite{PCAloss}, and our own extension of PCC, called Ensemble PCC, developed to enhance detection robustness through an ensemble of classifiers while preserving the simplicity of the original method.
In all the three algorithms, PCA-extracted features are combined with the input sample in a binary classifier to distinguish between normal and anomalous network traffic.
We use these algorithms on three standard network intrusion detection datasets: KDDCUP99, CICIDS2017, and DARKNET, with the goal of classifying normal versus anomalous traffic.

The machine learning training pipeline for these anomaly detection methods typically consists of three main stages: data preprocessing, PCA model extraction, and fine-tuning. 
Among these, the PCA model extraction step is often the most computationally intensive and resource-demanding, particularly in the classical setting. 
Our analysis specifically focuses on comparing the quantum and classical computational costs for this step, as it represents the primary computational bottleneck in the classical pipeline. 
However, classical preprocessing and fine-tuning can sometimes incur higher costs than the quantum model extraction step, highlighting the need for equivalent quantum algorithms to address these stages. 
While a thorough evaluation of quantum preprocessing and fine-tuning would be a valuable extension of this work, it lies beyond the scope of the present study.

We selected these PCA-based methods due to their simplicity and wide applicability.
First, simple QML algorithms are essential building blocks for developing more complex quantum models. 
Second, simple ML algorithms are particularly valuable in cybersecurity domains — such as defense and cyber-physical systems — where interpretability and explainability are crucial~\cite{geer2018rubicon}. 
These fields require transparency in automated decision-making processes, making simple, explainable models ideal for our evaluation.

Importantly, this paper does not aim to improve upon the latest state-of-the-art results in network intrusion detection. 
While more complex machine learning models may offer better detection performance, our goal is is to demonstrate the application of our evaluation framework in investigating quantum speedups. 
The simplicity of the algorithms chosen is deliberate, as it allows us to focus on the core aim of assessing quantum versus classical computational efficiency. 
This section showcases how to apply our methodology’s core steps (1, 2, 3, and Decision), with additional discussion on the resource estimation provided in Section~\ref{ssec:resest}.
Although future work may explore quantum speedups for more sophisticated models, such as neural networks or more advanced ensemble methods, these comparisons are beyond the scope of this study.

\begin{figure*}[ht]
     \centering
     \begin{subfigure}[b]{0.8\textwidth}
         \centering
         \includegraphics[width=\textwidth]{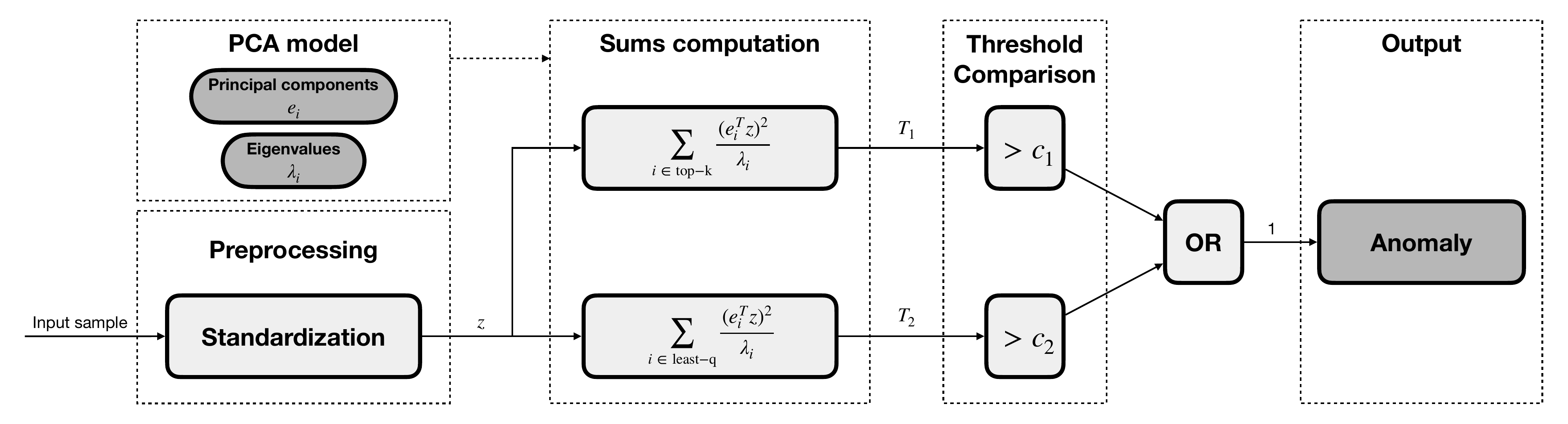}
    \caption{Principal Component Classifier (PCC).}
    \label{fig:pcc classifier}
     \end{subfigure}
     \hfill
    \centering
     \begin{subfigure}[b]{0.8\textwidth}
         \centering
         \includegraphics[width=\textwidth]{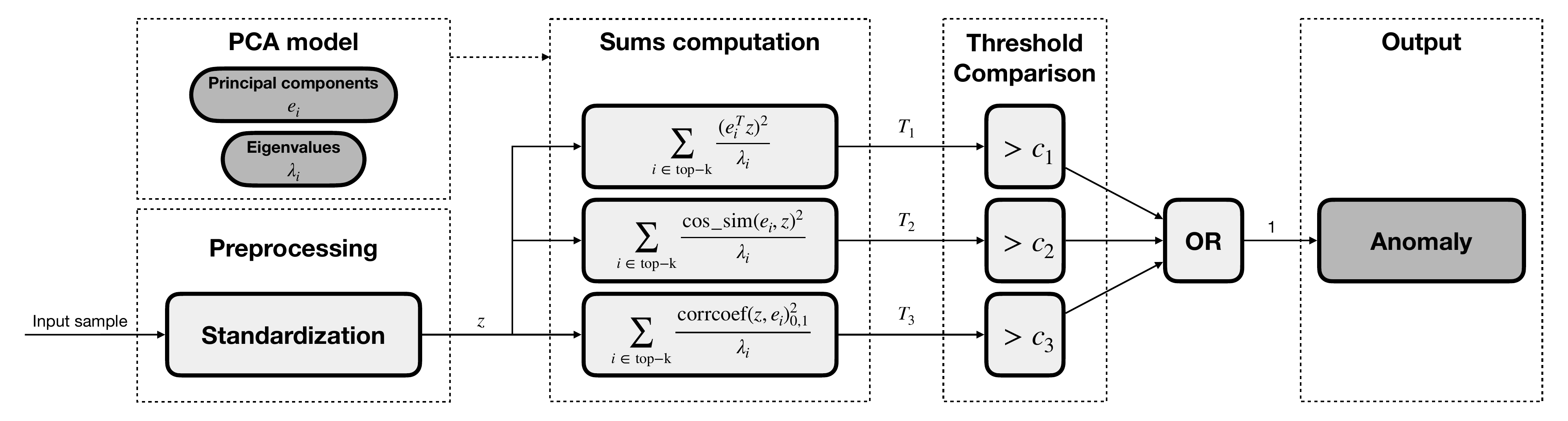}
    \caption{Ensemble PCC (major components).}
    \label{fig:ensemble classifier}
     \end{subfigure}
     \hfill
    \centering
     \begin{subfigure}[b]{0.8\textwidth}
         \centering
         \includegraphics[width=\textwidth]{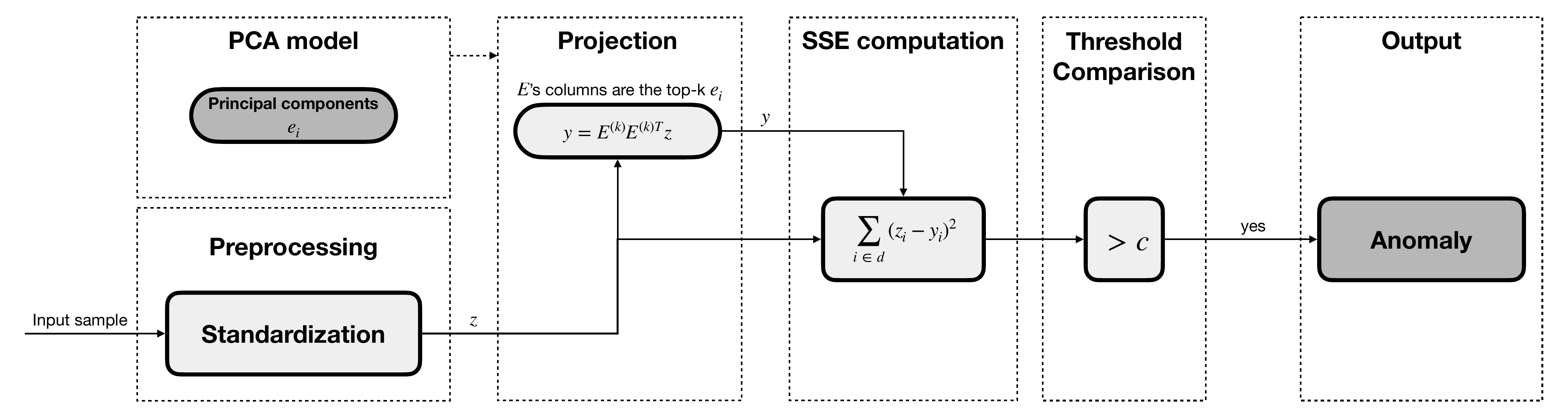}
    \caption{Reconstruction Loss.}
    \label{fig:RL classifier}
     \end{subfigure}
     \caption{Detection procedures of the three considered PCA-based anomaly detection classifiers.
     }
\end{figure*}

\begin{figure*}[ht]
     \centering
     \begin{subfigure}[b]{0.8\textwidth}
         \centering
         \includegraphics[width=\textwidth]{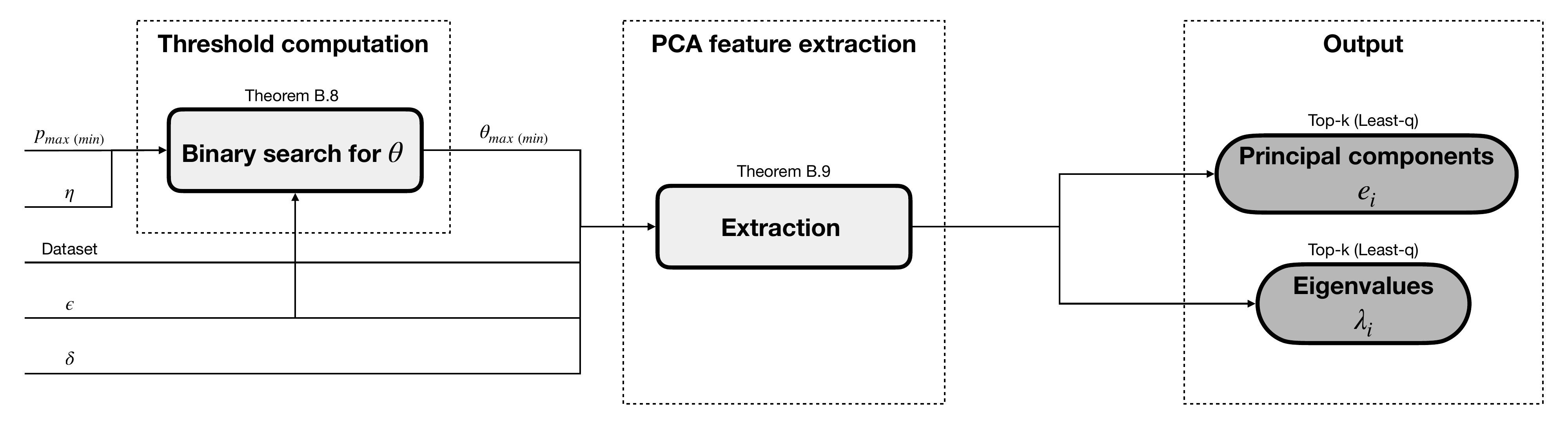}
     \end{subfigure}
    \caption{Quantum PCA model extraction.}
    \label{fig:quantum training}
\end{figure*}

\subsection{Anomaly detection algorithms}\label{ssec:anomaliydet}
In the following, we denote the standardized input data matrix as $\boldsymbol{X}\in\mathbb{R}^{n\times d}$, having $n$ samples, $d$ features, and rank $r$.
The principal components of $\boldsymbol{X}$ are the eigenvectors of the covariance matrix $\boldsymbol{X}^T\boldsymbol{X} \in \mathbb{R}^{d \times d}$, denoted by $\{\boldsymbol{e}_i\}_{i}^r$.
Their index always corresponds to the one of the eigenvalues $\{\lambda_i\}_i^{r}$, ordered decreasingly $\lambda_1 \geq \dots \geq \lambda_r$.
Given a set $S \in [r]$, the variance explained by the components in $S$ is $p = \frac{\sum_{i \in S} \lambda_i}{\sum_{j=1}^r \lambda_j} \in [0,1]$.

Fitting a PCA model means retrieving the principal components and corresponding eigenvalues that explain a total variance $p$. 
Usually, we are interested in the top-$k$ components (major), though sometimes the least-$q$ (minor) are of interest too.
In this case, we denote the largest threshold such that the principal components with eigenvalues $\sqrt{\lambda_i} > \theta$ explain variance $p_{maj}$ as $\theta \in (0,\sqrt{\lambda_1})$ and the smallest threshold such that the principal components with eigenvalues $\sqrt{\lambda_i} < \theta_{\min}$ explain variance $p_{min}$ as $\theta_{\min} \in (0,\sqrt{\lambda_1})$.
These parameters are summarized in Table~\ref{table:pca model parameters}.

We proceed to describe the three anomaly detection algorithms and report the quantum routines that allow model fitting.
The case study focuses on the advantages that the quantum routines can provide for the extracting the PCA model during the training phase, as this constitutes the algorithms' computational bottleneck. 
In all of these models, the PCA features are computed on a training set of normal (non-anomalous) data.

\begin{table}
\centering
\caption{PCA model parameters.}
\label{table:pca model parameters}
\begin{tabular}{|c|c|} 
    \hline
    Parameters & Description\\
    \hline
    $\boldsymbol{e}_i$ & $i^\mathrm{th}$ Principal component\\
    \hline
    $\lambda_i$ & \makecell{$i^\mathrm{th}$ Eigenvalue\\of the covariance matrix}\\
    \hline
    $p_{maj~(\min)}$ & \makecell{Amount of variance explained\\by the top (least) eigenvalues}\\
    \hline
    $\theta_{maj~(\min)}$ & \makecell{Cut-off threshold for\\the top (least) eigenvalues}\\
    \hline
    $k$ & Number of top eigenvalues\\
    \hline
    $q$ & Number of least eigenvalues\\
    \hline
\end{tabular}
\end{table}
\mypar{Principal Component Classifier (PCC)} 
The first anomaly detection algorithm is by Shyu et al.~\cite{PcaMajorMinor}. 
Combining the input sample and the extracted PCA features, the algorithm computes two sums and compares them against two anomaly detection threshold.
The two sums are $T_1=\sum_{i=1}^k\frac{y_i^2}{\lambda_i}$ and $\quad T_2=\sum_{i=r-q+1}^r\frac{y_i^2}{\lambda_i}$, where $k$ and $q$ are the number of major and minor principal components (explaining variance $p_{maj}$ and $p_{min}$) and $y_i=\boldsymbol{e}_i^T\boldsymbol{z}$, with $\boldsymbol{z}$ vector of standardized observations and $\boldsymbol{e_i}$ $i$-th eigenvector corresponding to the $\lambda_i$ eigenvalue. 
The algorithm classifies an observation $\boldsymbol{z}$ as an attack if $(T_1>c_1  \textbf{ or } T_2>c_2)$, where $c_1$ and $c_2$ are the outlier thresholds.
The outlier threshold are computed on a validation set using a function of $T_1$ and $T_2$, respectively, plus a parameter called false alarm rate $\alpha \in (0,1)$~\cite{PcaMajorMinor}. 
An increase of $\alpha$ corresponds to a decrease in the outlier thresholds, leading the algorithm to detect more anomalies at the expenses of false positives. 
Sometimes we only use the sum on the top principal components, without computing $T_2$ and $c_2$, we call this PCC with major components.
We summarize the PCC detection procedure in Figure~\ref{fig:pcc classifier}.

\mypar{Ensemble PCC}
We extend the PCC algorithm by proposing two novel ways of computing $y_i$, using cosine similarity and correlation measures, in addition to the original dot product between $\boldsymbol{e}_i$ and $\boldsymbol{z}$.
The cosine similarity between $\boldsymbol{e}_i$ and $\boldsymbol{z}$ is computed as $y_i = \frac{\boldsymbol{e}_i^T\boldsymbol{z}}{\|\boldsymbol{e}_i\|\|\boldsymbol{z}\|}$, while the correlation is computed using the $\mathrm{corrcoef}$ function from Numpy~\cite{harris2020array}.
Ensemble PCC computes three sums for the top-$k$ principal components and three for the least-$q$. 
Like in PCC, each sum is compared again a threshold. 
If any sum exceeds this threshold, then the input sample is labeled as an anomaly.
We call this method Ensemble PCC because the algorithm is equivalent to running three variations of PCC and labeling the sample as anomalous if any of the three models outputs so. 
We observe that this ensemble improves the performance of PCC.
Figure~\ref{fig:ensemble classifier} shows the Ensable PCC detection pipeline with major components.

\mypar{Reconstruction loss}
The reconstruction loss anomaly detection algorithm has been widely employed for a large variety of tasks. 
One example of its use in network intrusion detection is given by Verkerken et al.~\cite{PCAloss}. 
The key idea in this algorithm is that the top PCA components extracted from a training set of normal data can be used to explain normal data, but introduce errors when used to express anomalous data. 
To classify an input sample, the algorithm projects it onto the space of principal components and then back into the original feature space. 
This process of projecting the sample into a lower dimensional space and back leads to a reconstruction error: the method computes the loss as the sum of square error (SSE) and use it as an \emph{anomaly score}. 
The SSE sum is compared against a threshold and if the error is too large the sample gets labeled as an anomaly.
The detection pipeline is depicted in Figure~\ref{fig:RL classifier}.

\mypar{PCA model extraction}
The aim of the case study is to compare the classical and quantum running times needed to extract the PCA features from the training set of normal data points.
These features can then be used in all the three detection algorithms explained above.

The quantum routines that can be used to fit the PCA models are explained in detail by Bellante et al.~\cite{qadra}. 
The important routines are reported in Theorems~\ref{Theorivisto:binarysearch}, \ref{TheoMio:top-k_sv_extraction} and summarized in the following list.
\begin{itemize}
    \item \emph{Quantum binary search} for $\theta$ (Theorem~\ref{Theorivisto:binarysearch}).
    Given an amount $p$ of target explained variance and a tolerance parameter $\eta$, the quantum routine finds a threshold $\theta$ for the eigenvalues such that the top-$k$ (least-$q$, with a minor fix) components selected by $\theta$ explain at least $\overline{p}$ variance, with $\norm{p - \overline{p}} \leq \eta$. 
    Because of the quantum phase estimation error, the routine also needs an error parameter $\epsilon$ to estimate the eigenvalues on which the threshold $\theta$ is learned.
    The routine runs in $\widetilde{O}(\frac{\mu(\m{A})\log(\mu(\m{A})/\epsilon)}{\epsilon\eta})$.
    \item \emph{Quantum PCA extraction} (Theorem~\ref{TheoMio:top-k_sv_extraction}, note $\boldsymbol{e}_i = \boldsymbol{v}_i$).
    Given a threshold $\theta$, the second routine enables \emph{extracting} the eigenvalues $\lambda_i$ and the top principal components $\boldsymbol{e}_i$.
    The first task can be done in $\Ot(\frac{\lVert \boldsymbol{X}\rVert\mu(\boldsymbol{X})k\log(k)}{\theta\sqrt{p}\epsilon})$ to error $\|\lambda_i - \overline{\lambda}_i\| \leq 2\epsilon\sqrt{\lambda_i}$, while the latter takes $\Ot(dk\frac{\lVert \boldsymbol{X}\rVert\mu(\boldsymbol{X})\log(k)\log(d)}{\theta\sqrt{p}\epsilon\delta^2})$ to estimate the top-$k$ principal components such that $\|\boldsymbol{e}_i - \overline{\boldsymbol{e}}_i\|_2 \leq \delta$.
    This theorem can be modified (flip the condition of Alg. 4, step 3~\cite{qadra}) to extract the least-$q$ principal components and eigenvalues. 
    Provided a threshold $\theta_{\min}$, it can find the corresponding minor components in time $\widetilde{O}(\frac{\theta_{\min}}{\sigma_{\min}}\frac{\mu(\boldsymbol{X})}{\epsilon}\frac{qd}{\sqrt{p_{\min}}})$ with guarantee $\|\boldsymbol{e}_i - \overline{\boldsymbol{e}}_i\|_2 \leq \delta$.
\end{itemize}

The quantum PCA model extraction pipeline and the input parameters are represented in Figure~\ref{fig:quantum training}.

Depending on whether we require both the major and minor components or only the major ones, we compare the quantum running times with one of two classical alternatives: either the full singular value decomposition (SVD) with a complexity of $O(\min(nd^2, n^2d))$, or, if only the major components are needed, a randomized PCA variant with a lower complexity of $O(ndk\log(k))$~\cite{halko2010finding}.
The running time parameters of these algorithms are summarized in Table~\ref{table:quantum time parameters}.

\begin{table}[ht]
\centering
\caption{PCA model computation - running time parameters.}
\label{table:quantum time parameters}
\begin{tabular}{|c|c|} 
    \hline
    Parameters & Description\\
    \hline
    $n$ & Number of training datapoints\\
    \hline
    $d$ & \makecell{Number of features\\or principal components' size}\\
    \hline
    $k$ $(q)$ & \makecell{Number of top (least) principal\\components/eigenvalues to extract}\\
    \hline
    $\|\boldsymbol{X}\|$ & Dataset's spectral norm\\
    \hline
    $\mu(\boldsymbol{X})$ & \makecell{Dataset's normalization parameter\\(see Def~\ref{def:mu})}\\
    \hline
    $p_{maj~(\min)}$ & \makecell{Amount of variance explained\\by the top (least) eigenvalues}\\
    \hline
    $\theta_{maj~(\min)}$ & \makecell{Cut-off threshold for\\the top (least) eigenvalues}\\
    \hline
    $\sigma_{\min}$ & \makecell{Smallest value of all $\sqrt{\lambda_i}$}\\
    \hline
    $\eta$ & \makecell{Max error allowed on $p_{maj~(\min)}$ }\\
    \hline
    $\epsilon$ & \makecell{Max error allowed on each $\sqrt{\lambda_i}$}\\
    \hline
    $\delta$ & \makecell{Max error allowed on each $\boldsymbol{e}_i$,\\ in $\ell_2$ norm}\\
    \hline
\end{tabular}
\end{table}

\subsection{Experimental settings}
To simulate the PCA model extraction and perform the training, we selected classical PCA-based methods implemented in scikit-learn~\cite{scikit}, which uses the numerical methods of LAPACK~\cite{lapack99}, and modified them to model the error of the quantum subroutines\footnote{\url{https://github.com/tommasofioravanti/sq-learn}}. 

For the classical simulation of the quantum errors, we implemented Theorem~\ref{TheoMio:top-k_sv_extraction} for top-$k$ singular vector extractor (and Theorem~\ref{thm:q-means} for clustering), which are based on quantum pure state tomography (Theorem~\ref{thm:tomography}), amplitude estimation (Theorem~\ref{thm:amplitude estimation}), and phase estimation (Theorem~\ref{thm:phase-estimation}). 
Theoretically, these QML routines show a running time advantage over their classical counterparts with high-dimensional data.

We fit our model on three publicly available datasets:  KDDCUP99\footnote{We are aware of selected datasets' limitations (especially for KDDCUP99~\cite{tissec,cisda}). We use such datasets only to fairly compare the performance of classical and quantum algorithms, with the goal of understanding the long-term impact of quantum machine learning.}~\cite{cisda}, CIC-IDS2017~\cite{cicids2017}, and DARKNET~\cite{darknet}. 
In \ref{sec:tomography_est}, we report additional experiments on the tomography subroutine on the CIC-MALMEM-2022~\cite{carrier2022detecting} dataset.
For each dataset, we measure the parameters that influence the algorithms' running time and find the range of features and samples that would enable quantum advantage. 
The goal of this use case analysis is to compare the performance of training models using quantum algorithms against their classical counterparts by:
\textcircled{1} studying the influence of the intrinsic error that affects quantum-machine-learning algorithms on the detection task and \textcircled{2} evaluating the expected quantum running time as the dataset grows. 

Considering the characteristics of the dataset under analysis (i.e., number of features, number of points, effective condition number, approximation errors), we study the trade-off between the quantum algorithms' detection performance and running time, aiming at finding the \quotes{working point} that matches classical performances while minimizing the quantum algorithm's execution time. 
In particular, we fix the error parameters and evaluate the theoretical running time varying the number of samples and features.
We then compare the running times of the classical and the quantum anomaly detection models to quantify the dataset dimensions needed to observe any quantum advantage, discussing which cybersecurity tasks may match such requirements.

In our analysis, we do not consider the time required to acquire the data in a classical (quantum readable) memory, as these procedures are to be performed only once when the data is received and need to be done in either case. 
In \ref{apx:q-means}, we conduct further experiments on clustering with PCA-based dimensionality reduction and the quantum version of k-means. 
For this problem, while using a cybersecurity-related dataset, we do not perform anomaly detection, but we compare the quantum and the classical algorithm on a clustering metric and show that the two algorithms offer similar performance.

\mypar{Tomography heuristics} \ref{sec:tomography_est} discusses an analysis of our realistic simulation of the tomography subroutine.
Tomography sample complexity scales as $O\left(\frac{d\log d}{\delta^2}\right)$, where $d$ is the vector size. 
Experiments suggest that decreasing the number of samples by a significant constant factor still produces a near $\delta$-accurate reconstruction of the state vector.
We witness cases where we can save a constant factor up to $\approx 10^3$ in sample complexity from for $\delta=0.03, d=55$ (expected $\approx 10^7$). 
This heuristic is used in the results of Section~\ref{ssec:pca_loss}.

\subsection{PCC over KDDCUP99}
\label{ssec:pcckdd}

We execute the principal component classifier over the KDDCUP99 dataset. While we perform a set of experiments varying the percentage of variance retained by the models, here we report the results of PCA70 and QPCA70, which retain 70\% of the variance in the major components. 

Regarding the dataset, we consider numerical features
and split the dataset into a training set of $5,000$ \textit{normal} samples and a test set of  $92,278$ \textit{normal} samples and $39,674$ \textit{attacks}.
The training set is extracted using trimming to remove outliers and systematic random sampling (see \ref{ssec:furtherexp}).
Features are normalized and scaled for both test and training sets, with constant features removed.

\mypar{Performance analysis with Major Components only} We classify a sample as an attack only if $T_1>c_1$ and normal otherwise (with $c_1$ varying according to $\alpha$).
Results for PCA70 and QPCA70 are reported in Table \ref{table:q_exp1}, varying the false alarm rate $\alpha \in (0,1)$. 
For the quantum experiment, we consider the following error parameters: 
we use the quantum binary search of Theorem~\ref{Theorivisto:binarysearch} to estimate $\theta$, with parameters $p=0.70$,
$\epsilon_{\theta}=1$ as $\epsilon$, such that $|\sigma_i-\overline{\sigma}_i|\leq\epsilon_{\theta}$, and $\eta=0.1$ such that $|p-\overline{p}_{\theta}|\leq\eta$;
we use Theorem~\ref{TheoMio:top-k_sv_extraction} to extract $\boldsymbol{e}_i$ and $\lambda_i$ with parameters 
$\epsilon=1$ to estimate singular values such that $|\lambda_i-\overline{\lambda}_i|\leq\epsilon$, and $\delta=0.1$ is the error we tolerate in estimating singular vectors such that $\|\boldsymbol{e}_i-\overline{\boldsymbol{e}}_i\|_2\leq\delta$.
With these error parameters, we match classical performances. 
As $\alpha$ increases, recall increases, and precision decreases. This is expected because increasing $\alpha$ increases the false alarm rate, resulting in a lower outlier threshold and more observations being classified as attacks. In \ref{sec:pcckddmajorminor}, we extend this experiment using the minor components, too.

\begin{table}
\centering
\caption{Comparison for classical \textcircled{c} (PCA70) and quantum \textcircled{q} (QPCA70) principal components classifier with major components only over KDDCUP99.}
\label{table:q_exp1}
\resizebox{\columnwidth}{!}{
\begin{tabular}{|c|cc|cc|cc|cc|} 
    \hline
$\alpha$ (\%) & \multicolumn{2}{c|}{Recall (\%)}
            & \multicolumn{2}{c|}{Precision (\%)}& \multicolumn{2}{c|}{F1-score (\%)}
            & \multicolumn{2}{c|}{Accuracy (\%)}
              \\
    \hline
      &   c  &   q  &   c  &   q &   c  &   q  &   c  &   q   \\
    \hline
1  &  93.14    & 92.84     &98.63     &98.68  &  95.81    & 95.67     &97.55    &97.47     \\
    
2&  93.19 &92.88       &98.18       &98.23 &  95.62 &95.48       &97.43       &97.35                      \\
   
4   &96.04       &95.75       &96.51       &96.57  &96.28       &96.15       &97.76       &97.69                           \\
    
6   &98.51       &98.12       &94.20       &92.21     &96.30       &96.12       &97.72       &97.62                  \\

8   &98.67       &98.36       &92.01       &92.07   &95.22       &95.11       &97.02       &96.69           \\
   
10   &99.44       &99.12       &90.05       &90.10   &94.51       &94.40       &96.53       &96.46                 \\
\hline

\end{tabular}}
\end{table}

\begin{figure*}[t]
     \centering
     \begin{subfigure}[b]{0.3\textwidth}
         \centering
         \includegraphics[width=\textwidth]{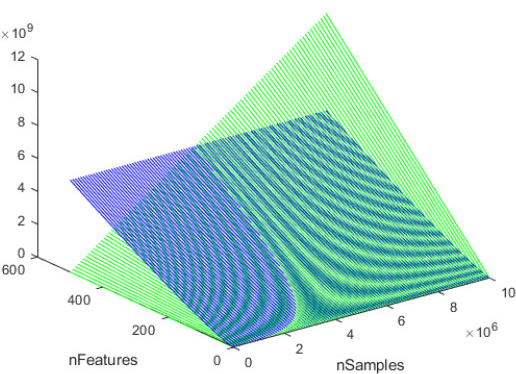}
    \caption{PCC major only on KDDCUP99.}
    \label{fig:pcc_runtime_comparisonKDD}
     \end{subfigure}
     \hfill
     \begin{subfigure}[b]{0.3\textwidth}
    \centering
    \includegraphics[width=\textwidth]{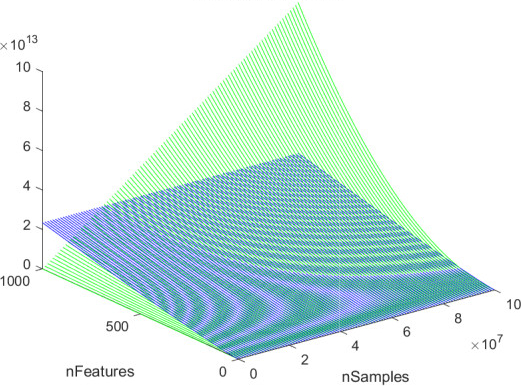}
    \caption{PCC/Ensamble on CICIDS2017}
    \label{fig:pcc_cicids_runtime}
     \end{subfigure}
     \hfill
    \begin{subfigure}[b]{0.3\textwidth}
    \centering
    \includegraphics[width=\textwidth]{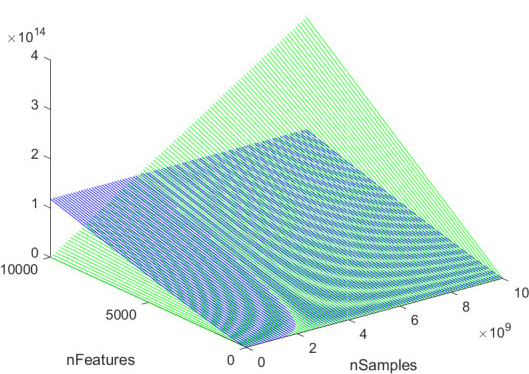}
    \caption{Rec. loss on CICIDS2017.}
    \label{fig:runtimePCAloss_}
    \end{subfigure}
     \caption{Running time comparison of classical (green) and quantum (blue) algorithms. In Plot~\ref{fig:pcc_runtime_comparisonKDD} we show the comparison of PCA and QPCA of Section~\ref{ssec:pcckdd} (KDDCUP99). 
     In Plot~\ref{fig:pcc_cicids_runtime} we show the running times of the algorithms discussed in Section~\ref{sec:pcc_mm_cicids} (CICIDS2017).
     In Plot~\ref{fig:runtimePCAloss_} we show the comparison between PCA and QPCA over the CICIDS2017 dataset,
     as per Section~\ref{ssec:pca_loss}. 
     }
\end{figure*}

\mypar{Running time analysis with Major Components only}
We compare the algorithms by plotting the classical and the quantum complexity, varying the number of samples $n$ and features $d$. 
For the quantum running time, we consider the time complexity of the quantum binary search and the quantum top-k singular vector extraction. 
The first has a cost of $\widetilde{O}\left(\frac{\mu(\boldsymbol{X})}{\epsilon\eta}\log(\frac{\mu(\boldsymbol{X})}{\epsilon})\right)$, while the latter requires 
$\Ot\left(dk\frac{\lVert \boldsymbol{X}\rVert\mu(\boldsymbol{X})}{\theta\sqrt{p}\epsilon\delta^2}\log(k)\log(d)\right)$ queries to estimate the top-k right singular vectors and values. 
We compare the quantum running time with a randomized classical version of PCA, which has a complexity of $O(ndk\log (k))$~\cite{halko2010finding} (since we are focusing on the major components only, there are better options than performing the full SVD, whose complexity is $O(\min\{nd^2,n^2d\})$).
Figure~\ref{fig:pcc_runtime_comparisonKDD} compares the quantum and classical running times in blue and green, respectively. 
We observe that the use of QML is not advantageous for small datasets. 
However, as the dataset's dimensionality increases, the query complexity advantage becomes evident (e.g., after $\approx 4*10^6$ samples and $\approx 50$ features).

\subsection{PCC and Ensemble over CICIDS2017}
\label{sec:pcc_mm_cicids}

We consider the principal component classifier with major and minor components and the CICIDS2017 dataset with DDoS attacks and normal samples. 
The training set comprises $5,000$ \textit{normal} samples, while the test set of $87,300$ \textit{normal} and $70,000$ \textit{DDoS} samples. 
We follow the same preprocessing performed on the KDDCUP dataset. 

\mypar{Performance analysis}
In Table~\ref{table:q_exp2}, we report the performance of QPCA70 (column q1) for PCC with major and minor components, using error parameters $\delta=0.1,\epsilon=\epsilon_{\theta}=1$, $\eta=0.1$, $\gamma=\frac{1}{d}$, $\theta_{\min}=\sqrt{0.20}$, and varying the false alarm rate $\alpha$. 
Using these parameters, we observe that the QPCA-70 model matches the performances of the classical equivalent (similar to the analysis of Table~\ref{table:q_exp1}).
The performances of both the classical and quantum PCC algorithms on CICIDS2017 (Table~\ref{table:q_exp2}, $q_1$) are significantly worse than the ones on KDDCUP99 with major components only (Table~\ref{table:q_exp1}).

Keeping the same parameters, we assess the detection performance of the ensemble method (an improvement of PCC, first proposed here).
The ensemble model improves the performance of PCC, particularly recall and accuracy, by utilizing six criteria for classifying an attack instead of two. 
This results in an improved recall, with a decrease in false negatives at the expense of false positives.
Despite the drop in precision, the substantial recall improvements account for an overall F1-score increase.
These results significantly improve over PCC, as seen in Table~\ref{table:q_exp2}.

\mypar{Running time analysis} 
We compare the running times by plotting the classical and the quantum complexity, varying the number of samples $n$ and features $d$. 
We consider the classical full-SVD time complexity $O\left(nd^2\right)$ to measure the classical running time, as we need all the components. 
For the quantum computation, instead, we consider the running time of quantum binary search and the quantum top-$k$ singular vectors extractor to estimate the top-$k$ components, plus the running time of the quantum least-$q$ singular vectors extractor to estimate the minor components, which is $\widetilde{O}\left(\frac{\theta_{\min}}{\sigma_{\min}}\frac{\mu(\boldsymbol{X})}{\epsilon}\frac{qd}{\sqrt{p_{\min}}}\right)$, where $\theta_{\min}$ is the custom $\theta$ passed to the function to extract the least singular values and $p_{\min}$ is the variance retained from the least components. 
In Figure~\ref{fig:pcc_cicids_runtime}, the quantum and the classical running times are presented in blue and green, respectively. 

As expected, the depicted running times are notably higher than those in the KDDCUP99 experiment, due to the comparison with classical full SVD complexity rather than the randomized one.
Additionally, the quantum case involves not only binary search and top-$k$ right singular vectors extraction but also the least-$q$ extraction, contributing to increased quantum running time. 
A quantum advantage in query complexity emerges with a large dataset of $\approx 2*10^{9}$ samples and $\approx 100$ features.

The absence of a quantum advantage over classical machine learning, as demonstrated by the full classical SVD with a dataset of $\approx 3*10^7$ samples and $500$ features, suggests that quantum is not particularly beneficial for problems requiring the extraction of minor components. 
Moreover, leveraging a more efficient classical algorithm for extracting minor components, such as Minor Component Analysis (MCA)~\cite{MCA}, would likely diminish the quantum advantage further, posing challenges for practical application in intrusion detection.

\begin{table}
\centering
\caption{Comparison for QPCA70 with both major and minor components ($q_1$) and QPCA70 ensemble ($q_2$) over CICIDS.}
\label{table:q_exp2}
\resizebox{\columnwidth}{!}{
\begin{tabular}{|c|cc|cc|cc|cc|}
    \hline
$\alpha$ (\%)  & \multicolumn{2}{c|}{Recall (\%)}
            & \multicolumn{2}{c|}{Precision (\%)}  & \multicolumn{2}{c|}{F1-score (\%)}
            & \multicolumn{2}{c|}{Accuracy (\%)}
              \\
    \hline
   &   $q_1$  &   $q_2$  &   $q_1$  &   $q_2$ &   $q_1$  &   $q_2$  &   $q_1$  &   $q_2$  \\
    \hline
1  &  36.05    & 39.80     &96.94     &95.30  &  52.55    & 56.15     &70.36    &71.69 \\
    
2&  58.97 &73.84       &96.54       &94.07 &  73.22 &82.74       &80.35       &85.97                     \\
   
4   &63.30       &89.61       &94.36      &90.79   &75.77      &90.19       &81.56       &91.13                 \\
    
6   &63.37      &96.96       &91.61       &87.25&74.92       &91.85       &80.68       &92.16                   \\

8   &64.43       &97.56       &88.99       &83.45&74.75       &89.95       &80.17       &90.08           \\
   
10   &65.90       &97.78       &86.92      &80.44 &74.97       &88.27       &79.96       &88.16                \\
\hline

\end{tabular}}
\end{table}

\subsection{Reconstruction loss over CICIDS2017}
\label{ssec:pca_loss}

We fit the PCA-based model with reconstruction loss over the CICIDS2017 dataset, including all the kinds of attacks in the anomalous class, following the same preprocessing of Section~\ref{sec:pcc_mm_cicids}. 
Unlike previous experiments, we perform hyper-parameter tuning on a validation set and preprocess the data with a quantile transform (see \ref{ssec:furtherexp}).
For the training set, we use $50,000$ normal samples; for the validation set, $60,000$ normal samples and $166,966$ attacks; finally, for the test set, we use $140,000$ normal samples and $389,590$ attacks, following~\cite{PCAloss}. 
Through hyper-parameters tuning, we found that the best PCA model has $12$ principal components, which retain $94.88\%$ of the variance. 
The outlier threshold is $t=0.425$: each sample whose anomaly score (defined in Section~\ref{ssec:anomaliydet}) is higher than $t$ is classified as anomalous.

\mypar{Performance analysis} We found that the error parameters needed to match the performances of the classical algorithm are $\epsilon_{\theta}=\epsilon=0.3$, $\eta=0.00075$, and we set $\gamma$ to $\frac{1}{d}$.
With these error parameters, we extract the same number of principal components of the classical model and explore perturbations on $\delta$ (the error on the singular vectors). 
Setting $\delta=0.01$, we match the classical performances. 
By increasing $\delta$, the model tends to classify data points as attacks, as shown by the precision and recall trends in Table~\ref{table:PCAReconstruction_delta}.

\begin{table}
\centering
\caption{Comparison of the classical (c) PCA and quantum (q) QPCA principal components classifier with reconstruction loss over CICIDS, varying $\delta$ error.}
\label{table:PCAReconstruction_delta}
\resizebox{\columnwidth}{!}{
\begin{tabular}{|c|cc|cc|cc|cc|}
\hline
$\delta$ &
  \multicolumn{2}{c|}{Recall (\%)} &
  \multicolumn{2}{c|}{Precision (\%)} &
  \multicolumn{2}{c|}{F1-score (\%)} &
  \multicolumn{2}{c|}{Accuracy (\%)} \\ \hline
 &
  \multicolumn{1}{c|}{q} &
  c &
  \multicolumn{1}{c|}{q} &
  c &
  \multicolumn{1}{c|}{q} &
  c &
  \multicolumn{1}{c|}{q} &
  c \\ \hline
0.01 &
  99.12 &
  \multirow{4}{*}{99.12} &
  91.28 &
  \multirow{4}{*}{91.28} &
  95.04 &
  \multirow{4}{*}{95.04} &
  98.08 &
  \multirow{4}{*}{98.08} \\
0.1 & 99.17 &  & 91.23 &  & 95.03 &  & 98.08 &  \\
0.9 & 99.79 &  & 71.31 &  & 83.18 &  & 92.52 &  \\
2   & 100   &  & 27.30 &  & 42.89 &  & 50.66 &  \\ \hline
\end{tabular}
}
\end{table}

\mypar{Running time analysis}
For this experiment, we used the heuristic insights obtained on the tomography subroutine (see Section~\ref{sec:case study} and \ref{sec:tomography_est}): we target an approximation error $\delta=0.01$ but divide the sample complexity of pure state tomography by a factor of $100$.
While this heuristic sacrifices provable guarantees, the model performs comparably to the classical counterpart in the testing phase.

Figure~\ref{fig:runtimePCAloss_} illustrates how the quantum and classical running times scale.
This model on CICIDS2017 has more principal components than PCC with major components over KDDCUP99, resulting in higher running times. 
Specifically, the current model requires $32$ components (retaining $p=99.75\%$ of the variance) compared to the PCC with major components only, which has $6$ components (retaining $p=70\%$ of the variance). 
For this model, the QML algorithms achieve advantage after $\approx 2 * 10^9$ samples.

\begin{figure}
    \centering
\includegraphics[width=0.3\textwidth]{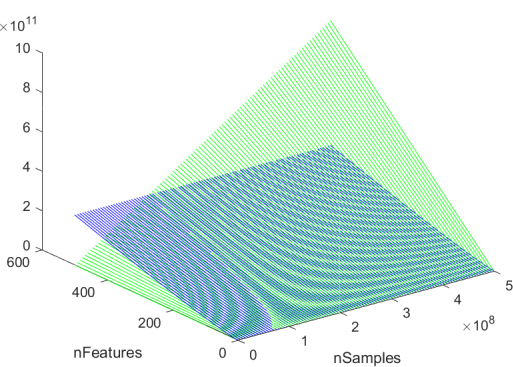}
\caption{Running time comparison of classical (green) and quantum (blue) algorithms for Rec. Loss over DARKNET.}
\label{fig:runtimePCAlossDARKNET}
\end{figure}

\subsection{Reconstruction loss over DARKNET}
\label{ssec:recolossdarknet}
We present the results obtained on PCA with Reconstruction loss over DARKNET.
The test set comprises $21,000$ normal samples and $14,000$ anomalies, the training set uses $50,000$ normal samples, and the validation set has $20,000$ normal samples plus $10,000$ anomalies. 
The $d=85$ features have been normalized and scaled to $0$ mean and unit variance. 
The hyperparameter search optimizing the F1-score over the classical PCA finds that the shorter running time is obtained with $35$ principal components (which retain $99.65\%$ of the variance) and outlier threshold $t=0.443$. 

\mypar{Performance analysis}
In the quantum model, the error parameters are set to $\epsilon_{\theta}=\epsilon=0.35,\eta=0.0011$, and $p=99.65\%$. 
With these parameters, $34$ principal components are extracted, aligning with the classical case. 
Table~\ref{table:LossDarknet_quantum} compares classical (c) and quantum (q) performances, showing how the model tends to classify data points as attacks as the error $\delta$ increases, similar to the previous CICIDS2017 experiment. 
We achieve almost the same performances of the classical algorithm by using the heuristic and setting the theoretical error on singular vectors to $\delta=0.1$ (i.e., setting $s=\frac{36d\log d}{\delta^2}$ with $\delta=0.01$ but taking $s/100$ samples).

\mypar{Running time analysis}
We plot the running time comparison in Figure~\ref{fig:runtimePCAlossDARKNET} using the parameters obtained from the hyperparameter search. 
We observe a quantum advantage in query complexity with $10^8$ samples for dataset features up to hundreds. 

\begin{table}
\centering
\caption{Comparison of the classical (c) PCA and quantum (q) QPCA principal components classifier with reconstruction loss over DARKNET, varying $\delta$ error. In the first row, we use the heuristic described in \ref{sec:tomography_est}. To get an error $\delta=0.01$ we should use $s=(36\frac{d\log d}{0.01})$ samples, but we see that $h=s/100$ suffice.}
\label{table:LossDarknet_quantum}
\resizebox{\columnwidth}{!}{
\begin{tabular}{|c|cc|cc|cc|cc|}
\hline
$\delta$ &
  \multicolumn{2}{c|}{Recall} &
  \multicolumn{2}{c|}{Precision} &
  \multicolumn{2}{c|}{F1-score} &
  \multicolumn{2}{c|}{Accuracy} \\ \hline
 &
  \multicolumn{1}{c|}{q} &
  c &
  \multicolumn{1}{c|}{q} &
  c &
  \multicolumn{1}{c|}{q} &
  c &
  \multicolumn{1}{c|}{q} &
  c \\ \hline
h &
  84.40 &
  \multirow{3}{*}{84.17} &
  73.18 &
  \multirow{3}{*}{73.71} &
  78.39 &
  \multirow{3}{*}{78.59} &
  81.07 &
  \multirow{3}{*}{81.35} \\
0.9 &
  87.08 &
   &
  68.12 &
   &
  76.44 &
   &
  78.17 &
   \\
2 &
  91.47 &
   &
  59.97 &
   &
  72.44 &
   &
  71.69 &
   \\ \hline
\end{tabular}
}
\end{table}

\subsection{Decision}
Upon initial examination, the dataset parameters required for the theoretical quantum advantage do not appear unrealistic, considering the volume of network packets companies receive daily. 
For instance, Microsoft's reported DDoS attack in January 2022 involved $3.47$ Terabit of data per second ($340$ millions of packets per second) \cite{MicrosoftDDoS}.
Using a training dataset from such an attack and considering $\approx 50$ features, the quantum model requires $\approx 1.3*10^8$ operations against $\approx3.9*10^{10}$ of the randomized classical model (see Figure~\ref{fig:pcc_runtime_comparisonKDD}).
Github also suffered a significant DDoS attack of about $1.35$ Terabit per second with about $130$ millions of packets per second~\cite{GitHubDDoS}. 
In this case, a quantum model would require $\approx1.4*10^8$ steps against $\approx1.3*10^{10}$ of the classical one.

While this suggests a theoretical gap in operations of about two orders of magnitude, the advantage appears only for massive datasets.
Therefore, the practical applicability of these QML algorithms for PCA-based network intrusion detection appears to be limited to large companies or organizations with the resources to handle vast amounts of data, computational power, time, and energy. 
In addition, for this particular PCA-based analysis, we witnessed high generalization performances by extracting the principal components on a small subset of data ($\approx 5000$ or $\approx 50000$ samples), rendering the need for such extensive datasets seemingly unrealistic for this application.

Despite these considerations, we proceed with a rough resource estimation to gauge the current state of quantum technology for these algorithms. 

\subsection{Resource estimation}\label{ssec:resest}
In this subsection, we analyze some key quantities for the oracle implementation of the QRAM, based on the code~\footnote{\url{https://github.com/glassnotes/FT_qRAM_Circuits/tree/master}} and analysis of Di Matteo et al.~\cite{di2020fault}. While we refrain from performing an exhaustive resource estimation, this preliminary analysis of the execution time of a QRAM query indicates the current state of quantum hardware. We focus on the QRAM \emph{Bucket Brigade parallel} circuit layout of Di Matteo et al.~\cite{di2020fault}, which provides reduced depth at the expenses of an increased number of auxiliary qubits. The analysis is performed on superconducting hardware with a defect-based error correcting surface code. 

We consider a dataset with $n=10^{ 7 }$ rows and $d=44$ features, stored in a KP-tree (see \ref{app:math}, near Theorem~\ref{thm:efficient data structure}) to allow quantum access to the dataset. This data structure consists of trees with a total of $O(nd\log(nd))$ nodes. The content of these nodes can be stored using an address space of $\lceil{\log_2(nm\log_2(nm))}\rceil = 34$ bits, assuming a system word size of 1 bit. Although a 1-bit word size is optimistic, it simplifies the circuit, allowing us to directly apply the architectural estimates from Di Matteo et al. In practice, a 32-bit word size might be more realistic, but even with the optimistic 1-bit assumption, the results demonstrate the substantial resources required by the QRAM circuit.

This configuration would require approximately $1.37 \times 10^{ 11 }$ logical qubits and a circuit depth of $539$ layers. This circuit also incurs a T-gate count of $3.61 \times 10^{ 11 }$, a T-depth of $67$, and a Clifford gate count of $9.28 \times 10^{ 11 }$. For error correction, we base our analysis on a superconducting architecture with a defect-based surface code. Considering the same parameters of Di Matteo et al., we assume hardware with a gate error probability of $10^{ -5 }$, a failure probability for the magic states in the first concatenation layer of the QECC of $10^{ -4 }$, and a surface code cycle of $200$ns.

While these assumptions are intentionally optimistic considering current hardware, they highlight the significant challenges that remain. Under these settings, a single QRAM query would take approximately $1.07$ms and require $2.08 \times 10^{ 14 }$ physical qubits. 
Looking more at the near future, less some realistic parameters would be hardware with a gate error probability of $10^{-3}$, a failure probability for the magic states in the first concatenation layer of the QECC of $10^{-2}$, and a surface code cycle of $1\mu$s. 
With these parameters, a single QRAM query would take approximately $28.1$ms and require $7.31 \times 10^{ 16 }$ physical qubits.

These numbers underscore the notable disparity in access time between a QRAM and a classical RAM, whose access time is in the scale of nanoseconds. This distinction in access time has cascading implications for the requirements on dataset size. Even a theoretical advantage of two orders of magnitude would not sufficiently compensate for the difference in memory access speed.

Finally, the sheer volume of physical qubits required is currently beyond the reach of today's technological trends. This underscores the practical limitations and emphasizes the considerable technological advancements needed to bridge the gap between theoretical potential and current quantum hardware capabilities for practical impacts.

\section{Conclusions}
\label{sec:conclusions}

In this work, we analyzed the potential impact of QML on cybersecurity tasks. 
We introduced a methodology for assessing quantum machine learning algorithms' impact on specific security problems. 
Besides laying out a clear methodology for studying quantum advantage in ML for cybersecurity, we demonstrated its application to fundamental algorithms used in network intrusion detection as a case study. 

The results presented in this work show that QML techniques are unlikely to outperform classical methods for intrusion detection on small datasets. 
In our case study, a quantum advantage in query complexity becomes apparent as the problem size surpasses $4 * 10^6$ samples and $50$ features.
There are settings where we observed significant theoretical gaps between the quantum and the classical models' running times. 
These requirements in the number of features and samples are potentially interesting for the intrusion detection domain since they align with publicly available datasets and real-world scenarios~\cite{MicrosoftDDoS,AWSDDoS,GitHubDDoS}. 
However, the hardware slowdowns highlighted in Sec.~\ref{ssec:resest} shift the advantage to bigger datasets, requiring even more physical qubits than estimated and making the advantage unlikely any soon.

Our case study suggests that fault-tolerant quantum computing could bring an asymptotic running time advantage, though more likely only for massive-scale datasets.  
However, the first generation of quantum computers is not expected to work on datasets of this size. 
Unless there is a significant advancement in hardware technology in the coming years, the sheer size of massive datasets will significantly exceed the capabilities of hardware implementations.

This evaluation framework can help the cybersecurity community find useful applications of quantum machine learning algorithms in practically relevant tasks and better distinguish the hype often associated with new technologies from their practical impact.
As more experts and practitioners from the field start practicing with the technologies, it will be possible to acquire a better understanding of the potential of quantum machine learning and of the way it will impact future users, markets, and attack-defense dynamics.

\section*{Acknowledgement} We thank Adithya Sireesh for useful discussions on error correction.
A.B. would like to thank professors Donatella Sciuto and Ferruccio Resta for their support and acknowledges the financial support from ICSC - “National Research Centre in High Performance Computing, Big Data and Quantum Computing” Spoke 10, funded by European Union – NextGenerationEU - under the grant PNRR-CN00000013-HPC.
Research at CQT is funded by the National Research Foundation, the Prime Minister’s Office, and the Ministry of Education, Singapore under the Research Centres of Excellence programme’s research grant R-710-000-012-135. We also acknowledge funding from the Quantum Engineering Programme (QEP 2.0) under grant NRF2021-QEP2-02-P05.


\begin{thebibliography}{100}

\bibitem{shor1994algorithms}
Peter~W Shor.
\newblock Algorithms for quantum computation: discrete logarithms and
  factoring.
\newblock In {\em Proceedings 35th annual symposium on foundations of computer
  science}, pages 124--134. Ieee, 1994.

\bibitem{bernstein2017post}
Daniel~J Bernstein and Tanja Lange.
\newblock Post-quantum cryptography.
\newblock {\em Nature}, 549(7671):188--194, 2017.

\bibitem{pirandola2020advances}
Stefano Pirandola, Ulrik~L Andersen, Leonardo Banchi, Mario Berta, Darius
  Bunandar, Roger Colbeck, Dirk Englund, Tobias Gehring, Cosmo Lupo, Carlo
  Ottaviani, et~al.
\newblock Advances in quantum cryptography.
\newblock {\em Advances in optics and photonics}, 12(4):1012--1236, 2020.

\bibitem{LLoyd}
Aram~W. Harrow, Avinatan Hassidim, and Seth Lloyd.
\newblock Quantum algorithm for linear systems of equations.
\newblock {\em Physical Review Letters}, 103(15), Oct 2009.

\bibitem{biamonte2017quantum}
Jacob Biamonte, Peter Wittek, Nicola Pancotti, Patrick Rebentrost, Nathan
  Wiebe, and Seth Lloyd.
\newblock Quantum machine learning.
\newblock {\em Nature}, 549(7671):195--202, 2017.

\bibitem{ucci2019survey}
Daniele Ucci, Leonardo Aniello, and Roberto Baldoni.
\newblock Survey of machine learning techniques for malware analysis.
\newblock {\em Computers \& Security}, 81:123--147, 2019.

\bibitem{rieck2011automatic}
Konrad Rieck, Philipp Trinius, Carsten Willems, and Thorsten Holz.
\newblock Automatic analysis of malware behavior using machine learning.
\newblock {\em Journal of computer security}, 19(4):639--668, 2011.

\bibitem{buczak2015survey}
Anna~L Buczak and Erhan Guven.
\newblock A survey of data mining and machine learning methods for cyber
  security intrusion detection.
\newblock {\em IEEE Communications surveys \& tutorials}, 18(2):1153--1176,
  2015.

\bibitem{Carminati2018}
Michele Carminati, Mario Polino, Andrea Continella, Andrea Lanzi, Federico
  Maggi, and Stefano Zanero.
\newblock Security evaluation of a banking fraud analysis system.
\newblock {\em ACM Trans. Priv. Secur.}, 21(3), April 2018.

\bibitem{arp2022and}
Daniel Arp, Erwin Quiring, Feargus Pendlebury, Alexander Warnecke, Fabio
  Pierazzi, Christian Wressnegger, Lorenzo Cavallaro, and Konrad Rieck.
\newblock Dos and don'ts of machine learning in computer security.
\newblock In {\em 31st USENIX Security Symposium (USENIX Security 22)}, pages
  3971--3988, 2022.

\bibitem{godefroid2017learn}
Patrice Godefroid, Hila Peleg, and Rishabh Singh.
\newblock Learn\&fuzz: Machine learning for input fuzzing.
\newblock In {\em 2017 32nd IEEE/ACM International Conference on Automated
  Software Engineering (ASE)}, pages 50--59. IEEE, 2017.

\bibitem{yamaguchi2011vulnerability}
Fabian Yamaguchi, Konrad Rieck, et~al.
\newblock Vulnerability extrapolation: Assisted discovery of vulnerabilities
  using machine learning.
\newblock In {\em 5th USENIX Workshop on Offensive Technologies (WOOT 11)},
  2011.

\bibitem{qmeans}
Iordanis Kerenidis, Jonas Landman, Alessandro Luongo, and Anupam Prakash.
\newblock q-means: {A} quantum algorithm for unsupervised machine learning.
\newblock In Hanna~M. Wallach, Hugo Larochelle, Alina Beygelzimer, Florence
  d'Alch{\'{e}}{-}Buc, Emily~B. Fox, and Roman Garnett, editors, {\em NeurIPS
  2019, December 8-14, 2019, Vancouver, BC, Canada}, 2019.

\bibitem{qadra}
Armando Bellante, Alessandro Luongo, and Stefano Zanero.
\newblock Quantum algorithms for svd-based data representation and analysis.
\newblock {\em Quantum Machine Intelligence}, 4(20), August 2022.

\bibitem{fraud_clust_1}
Andrei~Sorin Sabau.
\newblock Survey of clustering based financial fraud detection research.
\newblock {\em Informatica Economica}, 16(1):110, 2012.

\bibitem{pca_traffic_anomaly_detection_1}
Daniela Brauckhoff, Kave Salamatian, and Martin May.
\newblock Applying pca for traffic anomaly detection: Problems and solutions.
\newblock In {\em IEEE INFOCOM 2009}, pages 2866--2870. IEEE, 2009.

\bibitem{pca_traffic_anomaly_detection_2}
Haakon Ringberg, Augustin Soule, Jennifer Rexford, and Christophe Diot.
\newblock Sensitivity of pca for traffic anomaly detection.
\newblock In {\em Proceedings of the 2007 ACM SIGMETRICS international
  conference on Measurement and modeling of computer systems}, pages 109--120,
  2007.

\bibitem{pca_network_intrusion_1}
Nasrin Sultana, Naveen Chilamkurti, Wei Peng, and Rabei Alhadad.
\newblock Survey on sdn based network intrusion detection system using machine
  learning approaches.
\newblock {\em Peer-to-Peer Networking and Applications}, 12:493--501, 2019.

\bibitem{pca_dim_red_network}
K~Keerthi Vasan and B~Surendiran.
\newblock Dimensionality reduction using principal component analysis for
  network intrusion detection.
\newblock {\em Perspectives in Science}, 8:510--512, 2016.

\bibitem{hier_clust_1}
Shi-Jinn Horng, Ming-Yang Su, Yuan-Hsin Chen, Tzong-Wann Kao, Rong-Jian Chen,
  Jui-Lin Lai, and Citra~Dwi Perkasa.
\newblock A novel intrusion detection system based on hierarchical clustering
  and support vector machines.
\newblock {\em Expert systems with Applications}, 38(1):306--313, 2011.

\bibitem{kmeans_clust_1}
Meng Jianliang, Shang Haikun, and Bian Ling.
\newblock The application on intrusion detection based on k-means cluster
  algorithm.
\newblock In {\em 2009 International Forum on Information Technology and
  Applications}, volume~1, pages 150--152. IEEE, 2009.

\bibitem{spectral_clust_1}
Tao Ma, Fen Wang, Jianjun Cheng, Yang Yu, and Xiaoyun Chen.
\newblock A hybrid spectral clustering and deep neural network ensemble
  algorithm for intrusion detection in sensor networks.
\newblock {\em Sensors}, 16(10):1701, 2016.

\bibitem{malw_clust_2}
Swathi Pai, Fabio~Di Troia, Corrado~Aaron Visaggio, Thomas~H Austin, and Mark
  Stamp.
\newblock Clustering for malware classification.
\newblock {\em Journal of Computer Virology and Hacking Techniques},
  13:95--107, 2017.

\bibitem{svm_mal_1}
Min Zhao, Fangbin Ge, Tao Zhang, and Zhijian Yuan.
\newblock Antimaldroid: An efficient svm-based malware detection framework for
  android.
\newblock In {\em Information Computing and Applications: Second International
  Conference, ICICA 2011, Qinhuangdao, China, October 28-31, 2011. Proceedings,
  Part I 2}, pages 158--166. Springer, 2011.

\bibitem{svm_nn_intr_1}
Srinivas Mukkamala, Guadalupe Janoski, and Andrew Sung.
\newblock Intrusion detection using neural networks and support vector
  machines.
\newblock In {\em Proceedings of the 2002 International Joint Conference on
  Neural Networks. IJCNN'02 (Cat. No. 02CH37290)}, volume~2, pages 1702--1707.
  IEEE, 2002.

\bibitem{nn_phis}
Rami~M Mohammad, Fadi Thabtah, and Lee McCluskey.
\newblock Predicting phishing websites based on self-structuring neural
  network.
\newblock {\em Neural Computing and Applications}, 25:443--458, 2014.

\bibitem{nn_vuln}
Fang Wu, Jigang Wang, Jiqiang Liu, and Wei Wang.
\newblock Vulnerability detection with deep learning.
\newblock In {\em 2017 3rd IEEE international conference on computer and
  communications (ICCC)}, pages 1298--1302. IEEE, 2017.

\bibitem{cnn_malw_1}
Mahmoud Kalash, Mrigank Rochan, Noman Mohammed, Neil~DB Bruce, Yang Wang, and
  Farkhund Iqbal.
\newblock Malware classification with deep convolutional neural networks.
\newblock In {\em 2018 9th IFIP international conference on new technologies,
  mobility and security (NTMS)}, pages 1--5. IEEE, 2018.

\bibitem{cnn_malw_2}
Danish Vasan, Mamoun Alazab, Sobia Wassan, Hamad Naeem, Babak Safaei, and Qin
  Zheng.
\newblock Imcfn: Image-based malware classification using fine-tuned
  convolutional neural network architecture.
\newblock {\em Computer Networks}, 171:107138, 2020.

\bibitem{bharti2022noisy}
Kishor Bharti, Alba Cervera-Lierta, Thi~Ha Kyaw, Tobias Haug, Sumner
  Alperin-Lea, Abhinav Anand, Matthias Degroote, Hermanni Heimonen, Jakob~S
  Kottmann, Tim Menke, et~al.
\newblock Noisy intermediate-scale quantum algorithms.
\newblock {\em Reviews of Modern Physics}, 94(1):015004, 2022.

\bibitem{kalinin2023security}
Maxim Kalinin and Vasiliy Krundyshev.
\newblock Security intrusion detection using quantum machine learning
  techniques.
\newblock {\em Journal of Computer Virology and Hacking Techniques},
  19(1):125--136, 2023.

\bibitem{suryotrisongko2022evaluating}
Hatma Suryotrisongko and Yasuo Musashi.
\newblock Evaluating hybrid quantum-classical deep learning for cybersecurity
  botnet dga detection.
\newblock {\em Procedia Computer Science}, 197:223--229, 2022.

\bibitem{payares2021quantum}
ED~Payares and Juan~Carlos Mart{\'\i}nez-Santos.
\newblock Quantum machine learning for intrusion detection of distributed
  denial of service attacks: a comparative overview.
\newblock {\em Quantum Computing, Communication, and Simulation}, 11699:35--43,
  2021.

\bibitem{masum2022quantum}
Mohammad Masum, Mohammad Nazim, Md~Jobair~Hossain Faruk, Hossain Shahriar,
  Maria Valero, Md~Abdullah~Hafiz Khan, Gias Uddin, Shabir Barzanjeh, Erhan
  Saglamyurek, Akond Rahman, et~al.
\newblock Quantum machine learning for software supply chain attacks: How far
  can we go?
\newblock In {\em 2022 IEEE 46th Annual Computers, Software, and Applications
  Conference (COMPSAC)}, pages 530--538. IEEE, 2022.

\bibitem{beaudoin2022quantum}
Collin Beaudoin, Satwik Kundu, Rasit~Onur Topaloglu, and Swaroop Ghosh.
\newblock Quantum machine learning for material synthesis and hardware
  security.
\newblock In {\em Proceedings of the 41st IEEE/ACM International Conference on
  Computer-Aided Design}, pages 1--7, 2022.

\bibitem{aharonov2023polynomial}
Dorit Aharonov, Xun Gao, Zeph Landau, Yunchao Liu, and Umesh Vazirani.
\newblock A polynomial-time classical algorithm for noisy random circuit
  sampling.
\newblock In {\em Proceedings of the 55th Annual ACM Symposium on Theory of
  Computing}, pages 945--957, 2023.

\bibitem{arthur2021qubo}
Davis Arthur, Lauren Pusey-Nazzaro, et~al.
\newblock Qubo formulations for training machine learning models.
\newblock {\em Scientific reports}, 11(1):1--10, 2021.

\bibitem{aimeur2013quantum}
Esma A{\"\i}meur, Gilles Brassard, and S{\'e}bastien Gambs.
\newblock Quantum speed-up for unsupervised learning.
\newblock {\em Machine Learning}, 90(2):261--287, 2013.

\bibitem{kerenidis2021quantumspectral}
Iordanis Kerenidis and Jonas Landman.
\newblock Quantum spectral clustering.
\newblock {\em Physical Review A}, 103(4):042415, 2021.

\bibitem{wiebe2015quantum}
Nathan Wiebe, Ashish Kapoor, and Krysta~M Svore.
\newblock Quantum nearest-neighbor algorithms for machine learning.
\newblock {\em Quantum information and computation}, 15(3-4):318--358, 2015.

\bibitem{lloyd2013quantum}
Seth Lloyd, Masoud Mohseni, and Patrick Rebentrost.
\newblock Quantum algorithms for supervised and unsupervised machine learning.
\newblock {\em arXiv preprint arXiv:1307.0411}, 2013.

\bibitem{bellante2023quantum}
Armando Bellante, William Bonvini, Stefano Vanerio, and Stefano Zanero.
\newblock Quantum eigenfaces: Linear feature mapping and nearest neighbor
  classification with outlier detection.
\newblock In {\em 2023 IEEE International Conference on Quantum Computing and
  Engineering (QCE)}, volume~1, pages 196--207. IEEE, 2023.

\bibitem{kerenidis2020quantumgaussian}
Iordanis Kerenidis, Alessandro Luongo, and Anupam Prakash.
\newblock Quantum expectation-maximization for gaussian mixture models.
\newblock In {\em International Conference on Machine Learning}, pages
  5187--5197. PMLR, 2020.

\bibitem{qmp}
Armando Bellante and Stefano Zanero.
\newblock Quantum matching pursuit: A quantum algorithm for sparse
  representations.
\newblock {\em Phys. Rev. A}, 105:022414, Feb 2022.

\bibitem{rebentrost2014quantum}
Patrick Rebentrost, Masoud Mohseni, and Seth Lloyd.
\newblock Quantum support vector machine for big data classification.
\newblock {\em Physical review letters}, 113(13):130503, 2014.

\bibitem{chakraborty2023quantum}
Shantanav Chakraborty, Aditya Morolia, and Anurudh Peduri.
\newblock Quantum regularized least squares.
\newblock {\em Quantum}, 7:988, 2023.

\bibitem{kerenidis2020classificationqsfa}
Iordanis Kerenidis and Alessandro Luongo.
\newblock Classification of the mnist data set with quantum slow feature
  analysis.
\newblock {\em Physical Review A}, 101(6):062327, 2020.

\bibitem{roget2022quantum}
Mathieu Roget, Giuseppe Di~Molfetta, and Hachem Kadri.
\newblock Quantum perceptron revisited: Computational-statistical tradeoffs.
\newblock In {\em Uncertainty in Artificial Intelligence}, pages 1697--1706.
  PMLR, 2022.

\bibitem{liao2021quadratic}
Pengcheng Liao, Barry~C Sanders, and Tim Byrnes.
\newblock Quadratic quantum speedup for perceptron training.
\newblock {\em arXiv preprint arXiv:2109.04695}, 2021.

\bibitem{kerenidis2019quantumQCNN}
Iordanis Kerenidis, Jonas Landman, and Anupam Prakash.
\newblock Quantum algorithms for deep convolutional neural networks.
\newblock In {\em 8th International Conference on Learning Representations,
  {ICLR} 2020, Addis Ababa, Ethiopia, April 26-30, 2020}, 2020.

\bibitem{dalzell2023end}
Alexander~M Dalzell, B~David Clader, Grant Salton, Mario Berta, Cedric Yen-Yu
  Lin, David~A Bader, Nikitas Stamatopoulos, Martin~JA Schuetz, Fernando~GSL
  Brand{\~a}o, Helmut~G Katzgraber, et~al.
\newblock End-to-end resource analysis for quantum interior-point methods and
  portfolio optimization.
\newblock {\em PRX Quantum}, 4(4):040325, 2023.

\bibitem{nielsen2010quantum}
Michael~A Nielsen and Isaac~L Chuang.
\newblock {\em Quantum Computation and Quantum Information}.
\newblock Cambridge University Press, 2010.

\bibitem{madsen2022quantum}
Lars~S Madsen, Fabian Laudenbach, Mohsen~Falamarzi Askarani, Fabien Rortais,
  Trevor Vincent, Jacob~FF Bulmer, Filippo~M Miatto, Leonhard Neuhaus, Lukas~G
  Helt, Matthew~J Collins, et~al.
\newblock Quantum computational advantage with a programmable photonic
  processor.
\newblock {\em Nature}, 606(7912):75--81, 2022.

\bibitem{preskill2012quantum}
John Preskill.
\newblock Quantum computing and the entanglement frontier.
\newblock {\em arXiv preprint arXiv:1203.5813}, 2012.

\bibitem{saffman2010quantum}
Mark Saffman, Thad~G Walker, and Klaus M{\o}lmer.
\newblock Quantum information with rydberg atoms.
\newblock {\em Reviews of modern physics}, 82(3):2313, 2010.

\bibitem{bruzewicz2019trapped}
Colin~D Bruzewicz, John Chiaverini, Robert McConnell, and Jeremy~M Sage.
\newblock Trapped-ion quantum computing: Progress and challenges.
\newblock {\em Applied Physics Reviews}, 6(2):021314, 2019.

\bibitem{kjaergaard2020superconducting}
Morten Kjaergaard, Mollie~E Schwartz, Jochen Braum{\"u}ller, Philip Krantz,
  Joel I-J Wang, Simon Gustavsson, and William~D Oliver.
\newblock Superconducting qubits: Current state of play.
\newblock {\em Annual Review of Condensed Matter Physics}, 11:369--395, 2020.

\bibitem{babbush2021focus}
Ryan Babbush, Jarrod~R McClean, Michael Newman, Craig Gidney, Sergio Boixo, and
  Hartmut Neven.
\newblock Focus beyond quadratic speedups for error-corrected quantum
  advantage.
\newblock {\em PRX Quantum}, 2(1):010103, 2021.

\bibitem{maronese2022quantum}
Marco Maronese, Lorenzo Moro, Lorenzo Rocutto, and Enrico Prati.
\newblock Quantum compiling.
\newblock In {\em Quantum Computing Environments}, pages 39--74. Springer,
  2022.

\bibitem{bouland2021efficient}
Adam Bouland and Tudor Giurgica-Tiron.
\newblock Efficient universal quantum compilation: An inverse-free
  solovay-kitaev algorithm.
\newblock {\em arXiv preprint arXiv:2112.02040}, 2021.

\bibitem{duncan2020graph}
Ross Duncan, Aleks Kissinger, Simon Perdrix, and John Van De~Wetering.
\newblock Graph-theoretic simplification of quantum circuits with the
  zx-calculus.
\newblock {\em Quantum}, 4:279, 2020.

\bibitem{ambainis2000quantum}
Andris Ambainis.
\newblock Quantum lower bounds by quantum arguments.
\newblock In {\em Proceedings of the thirty-second annual ACM symposium on
  Theory of computing}, pages 636--643, 2000.

\bibitem{beals2001quantum}
Robert Beals, Harry Buhrman, Richard Cleve, Michele Mosca, and Ronald De~Wolf.
\newblock Quantum lower bounds by polynomials.
\newblock {\em Journal of the ACM (JACM)}, 48(4):778--797, 2001.

\bibitem{grover1997quantum}
Lov~K Grover.
\newblock Quantum mechanics helps in searching for a needle in a haystack.
\newblock {\em Physical review letters}, 79(2):325, 1997.

\bibitem{amy2016estimating}
Matthew Amy, Olivia~Di Matteo, Vlad Gheorghiu, Michele Mosca, Alex Parent, and
  John Schanck.
\newblock Estimating the cost of generic quantum pre-image attacks on sha-2 and
  sha-3.
\newblock In {\em International Conference on Selected Areas in Cryptography},
  pages 317--337. Springer, 2016.

\bibitem{aggarwal2018quantum}
Divesh Aggarwal, Gavin Brennen, Troy Lee, Miklos Santha, and Marco Tomamichel.
\newblock Quantum attacks on bitcoin, and how to protect against them.
\newblock {\em Ledger}, 3, 2018.

\bibitem{ettinger2004quantum}
Mark Ettinger, Peter H{\o}yer, and Emanuel Knill.
\newblock The quantum query complexity of the hidden subgroup problem is
  polynomial.
\newblock {\em Information Processing Letters}, 91(1):43--48, 2004.

\bibitem{low2019hamiltonian}
Guang~Hao Low and Isaac~L Chuang.
\newblock Hamiltonian simulation by qubitization.
\newblock {\em Quantum}, 3:163, 2019.

\bibitem{camps2024explicit}
Daan Camps, Lin Lin, Roel Van~Beeumen, and Chao Yang.
\newblock Explicit quantum circuits for block encodings of certain sparse
  matrices.
\newblock {\em SIAM Journal on Matrix Analysis and Applications},
  45(1):801--827, 2024.

\bibitem{di2020fault}
Olivia Di~Matteo, Vlad Gheorghiu, and Michele Mosca.
\newblock Fault-tolerant resource estimation of quantum random-access memories.
\newblock {\em IEEE Transactions on Quantum Engineering}, 1:1--13, 2020.

\bibitem{giovannetti2008architectures}
Vittorio Giovannetti, Seth Lloyd, and Lorenzo Maccone.
\newblock Architectures for a quantum random access memory.
\newblock {\em Physical Review A}, 78(5):052310, 2008.

\bibitem{hann2021resilience}
Connor~T Hann, Gideon Lee, SM~Girvin, and Liang Jiang.
\newblock Resilience of quantum random access memory to generic noise.
\newblock {\em PRX Quantum}, 2(2):020311, 2021.

\bibitem{hann2019hardware}
Connor~T Hann, Chang-Ling Zou, Yaxing Zhang, Yiwen Chu, Robert~J Schoelkopf,
  Steven~M Girvin, and Liang Jiang.
\newblock Hardware-efficient quantum random access memory with hybrid quantum
  acoustic systems.
\newblock {\em Physical Review Letters}, 123(25):250501, 2019.

\bibitem{liu2022quantum}
Junyu Liu, Connor~T Hann, and Liang Jiang.
\newblock Quantum data center: Theories and applications.
\newblock {\em arXiv preprint arXiv:2207.14336}, 2022.

\bibitem{van2023quantum}
Joran van Apeldoorn, Arjan Cornelissen, Andr{\'a}s Gily{\'e}n, and Giacomo
  Nannicini.
\newblock Quantum tomography using state-preparation unitaries.
\newblock In {\em Proceedings of the 2023 Annual ACM-SIAM Symposium on Discrete
  Algorithms (SODA)}, pages 1265--1318. SIAM, 2023.

\bibitem{tomography}
Iordanis Kerenidis and Anupam Prakash.
\newblock A quantum interior point method for lps and sdps.
\newblock {\em ACM Transactions on Quantum Computing}, 1(1):1--32, 2020.

\bibitem{aaronson2015read}
Scott Aaronson.
\newblock Read the fine print.
\newblock {\em Nature Physics}, 11(4):291--293, 2015.

\bibitem{PcaMajorMinor}
Mei-Ling Shyu, Shu-Ching Chen, Kanoksri Sarinnapakorn, and Liwu Chang.
\newblock A novel anomaly detection scheme based on principal component
  classifier.
\newblock In {\em in Proceedings of the IEEE Foundations and New Directions of
  Data Mining Workshop, in conjunction with the Third IEEE International
  Conference on Data Mining (ICDM’03)}, 01 2003.

\bibitem{PCAloss}
Miel Verkerken, Laurens D’hooge, Tim Wauters, Bruno Volckaert, and Filip
  De~Turck.
\newblock Unsupervised machine learning techniques for network intrusion
  detection on modern data.
\newblock In {\em 2020 4th Cyber Security in Networking Conference (CSNet)},
  pages 1--8, 2020.

\bibitem{geer2018rubicon}
Daniel Earl Geer~Jr.
\newblock A rubicon.
\newblock {\em Aegis Series Paper No. 1801}, 2018.

\bibitem{harris2020array}
Charles~R. Harris, K.~Jarrod Millman, St{\'{e}}fan~J. van~der Walt, Ralf
  Gommers, Pauli Virtanen, David Cournapeau, Eric Wieser, Julian Taylor,
  Sebastian Berg, Nathaniel~J. Smith, Robert Kern, Matti Picus, Stephan Hoyer,
  Marten~H. van Kerkwijk, Matthew Brett, Allan Haldane, Jaime~Fern{\'{a}}ndez
  del R{\'{i}}o, Mark Wiebe, Pearu Peterson, Pierre G{\'{e}}rard-Marchant,
  Kevin Sheppard, Tyler Reddy, Warren Weckesser, Hameer Abbasi, Christoph
  Gohlke, and Travis~E. Oliphant.
\newblock Array programming with {NumPy}.
\newblock {\em Nature}, 585(7825):357--362, September 2020.

\bibitem{halko2010finding}
Nathan Halko, Per-Gunnar Martinsson, and Joel~A. Tropp.
\newblock Finding structure with randomness: Probabilistic algorithms for
  constructing approximate matrix decompositions.
\newblock {\em SIAM Rev., Survey and Review section, Vol. 53, num. 2, pp.
  217-288, June 2011}, 2009.

\bibitem{scikit}
F.~Pedregosa, G.~Varoquaux, A.~Gramfort, V.~Michel, B.~Thirion, O.~Grisel,
  M.~Blondel, P.~Prettenhofer, R.~Weiss, V.~Dubourg, J.~Vanderplas, A.~Passos,
  D.~Cournapeau, M.~Brucher, M.~Perrot, and E.~Duchesnay.
\newblock Scikit-learn: Machine learning in {P}ython.
\newblock {\em Journal of Machine Learning Research}, 12:2825--2830, 2011.

\bibitem{lapack99}
E.~Anderson, Z.~Bai, C.~Bischof, S.~Blackford, J.~Demmel, J.~Dongarra,
  J.~Du~Croz, A.~Greenbaum, S.~Hammarling, A.~McKenney, and D.~Sorensen.
\newblock {\em {LAPACK} Users' Guide}.
\newblock Society for Industrial and Applied Mathematics, Philadelphia, PA,
  third edition, 1999.

\bibitem{tissec}
John McHugh.
\newblock Testing intrusion detection systems: a critique of the 1998 and 1999
  {DARPA} intrusion detection system evaluations as performed by lincoln
  laboratory.
\newblock {\em {ACM} Trans. Inf. Syst. Secur.}, 3(4):262--294, 2000.

\bibitem{cisda}
Mahbod Tavallaee, Ebrahim Bagheri, Wei Lu, and Ali~A. Ghorbani.
\newblock A detailed analysis of the {KDD} {CUP} 99 data set.
\newblock In {\em 2009 {IEEE} Symposium on Computational Intelligence for
  Security and Defense Applications, {CISDA} 2009, Ottawa, Canada, July 8-10,
  2009}, pages 1--6. {IEEE}, 2009.

\bibitem{cicids2017}
Iman Sharafaldin, Arash Habibi~Lashkari, and Ali Ghorbani.
\newblock Toward generating a new intrusion detection dataset and intrusion
  traffic characterization.
\newblock In {\em ICISSP}, pages 108--116, 01 2018.

\bibitem{darknet}
Arash Habibi~Lashkari, Gurdip Kaur, and Abir Rahali.
\newblock Didarknet: A contemporary approach to detect and characterize the
  darknet traffic using deep image learning.
\newblock In {\em 2020 the 10th International Conference on Communication and
  Network Security}, ICCNS 2020, page 1–13, New York, NY, USA, 2020.
  Association for Computing Machinery.

\bibitem{carrier2022detecting}
Tristan Carrier, Princy Victor, Ali Tekeoglu, and Arash~Habibi Lashkari.
\newblock Detecting obfuscated malware using memory feature engineering.
\newblock In {\em ICISSP}, pages 177--188, 2022.

\bibitem{MCA}
Fa-Long Luo, Rolf Unbehauen, and Andrzej Cichocki.
\newblock A minor component analysis algorithm.
\newblock {\em Neural Networks}, 10(2):291--297, 1997.

\bibitem{MicrosoftDDoS}
Alethea Toh.
\newblock Azure ddos protection—2021 q3 and q4 ddos attack trends, 2021.
\newblock Archived from the original on 02/02/2022:
  https://web.archive.org/web/20230203101146/https://a
  zure.microsoft.com/en-us/blog/azure-ddos-protection-2021-q3-and-q4-ddos-attack-trends/.

\bibitem{GitHubDDoS}
Sam Kottler.
\newblock Github ddos attack in 2018, 2018.
\newblock Archived from the original on 02/02/2022:
  https://web.archive.org/web/20230203102859/https://git
  hub.blog/2018-03-01-ddos-incident-report/.

\bibitem{AWSDDoS}
BBC.
\newblock Amazon 'thwarts largest ever ddos cyber-attack', 2020.
\newblock Archived from the original on 02/02/2022:
  https://web.archive.org/save/https://www.bbc.com/news/ technology-53093611.

\bibitem{kerenidis2020quantumGRADIENT}
Iordanis Kerenidis and Anupam Prakash.
\newblock Quantum gradient descent for linear systems and least squares.
\newblock {\em Physical Review A}, 101(2):022316, 2020.

\bibitem{kerenidis2016quantumREC}
Iordanis Kerenidis and Anupam Prakash.
\newblock Quantum recommendation systems.
\newblock In Christos~H. Papadimitriou, editor, {\em 8th Innovations in
  Theoretical Computer Science Conference, {ITCS} 2017, January 9-11, 2017,
  Berkeley, CA, {USA}}, volume~67 of {\em LIPIcs}, pages 49:1--49:21. Schloss
  Dagstuhl - Leibniz-Zentrum f{\"{u}}r Informatik, 2017.

\bibitem{kerenidis2020quantumIP}
Iordanis Kerenidis and Anupam Prakash.
\newblock A quantum interior point method for lps and sdps.
\newblock {\em ACM Transactions on Quantum Computing}, 1(1):1--32, 2020.

\bibitem{ta2013inverting}
Amnon Ta-Shma.
\newblock Inverting well conditioned matrices in quantum logspace.
\newblock In {\em Proceedings of the forty-fifth annual ACM symposium on Theory
  of computing}, pages 881--890, 2013.

\bibitem{brassard2002quantum}
Gilles Brassard, Peter Hoyer, Michele Mosca, and Alain Tapp.
\newblock Quantum amplitude amplification and estimation.
\newblock {\em Contemporary Mathematics}, 305:53--74, 2002.

\bibitem{tommythesis}
Tommaso Fioravanti.
\newblock Evaluation of quantum machine learning algorithms for cybersecurity.
\newblock Master's thesis, Politecnico di Milano, 2022.

\bibitem{wrapperapproach}
Ron Kohavi and George~H. John.
\newblock Wrappers for feature subset selection.
\newblock {\em Artif. Intell.}, 97(1-2):273--324, 1997.

\bibitem{optuna_2019}
Takuya Akiba, Shotaro Sano, Toshihiko Yanase, Takeru Ohta, and Masanori Koyama.
\newblock Optuna: A next-generation hyperparameter optimization framework.
\newblock In {\em Proceedings of the 25rd {ACM} {SIGKDD} International
  Conference on Knowledge Discovery and Data Mining}, 2019.

\end{thebibliography}

\appendix

\section{Brief introduction to quantum information}
\label{apx:intro quantum}

The fundamental unit of quantum information is the qubit.
The state of a qubit can be expressed as a linear combination of vectors from an orthonormal basis of $\C^2$, such that the sum of squares of the absolute values of the coefficients sums up to one.
Often, we express states using the computational basis $\ket{0} = \begin{bmatrix}
    1 & 0
\end{bmatrix}^T$, $\ket{1} = \begin{bmatrix}
    0 & 1
\end{bmatrix}^T$, meaning that a generic qubit state can be written as $\ket{\varphi} = \alpha\ket{0} + \beta\ket{1}$, with $\alpha, \beta \in \C$ and $\abs{\alpha}^2 + \abs{\beta}^2 = 1$.
The coefficients are also called \emph{amplitudes} and if more than one is non-zero, we say that the state is in a \emph{superposition} of the basis states. 

A quantum register is an ensemble of qubits.
We can express the state of a quantum register as $\ket{v} = \frac{1}{\norm{v}}\sum_{i=0}^{2^n-1} v_i \ket{i} \in \C^{2^n}$, where $n\in \N$ is the number of qubits in the register, each $v_i \in \C$, and $\sum_{i=0}^{2^n-1}\abs{v_i}^2 = 1$.
Here $\ket{i}$ is the $i$-th vector of the computational basis of $\C^{2^n}$; i.e., $\ket{i} = \ket{i_0}\otimes\ket{i_1}\otimes\dots \otimes\ket{i_{n-1}}$, where $i_j$ denotes the $j^{\mathrm{th}}$ bit of the binary encoding of $i$. 
This representation is analogous to the decomposition of a vector in the computational basis of a vector space. 
It is crucial to note that $n$ qubits suffice to span a space of dimension $2^n$. 

A quantum algorithm consists of \emph{evolving} and \emph{measuring} a certain initial state, multiple times. 

\noindent a) The \emph{evolution} of a $n$-qubits quantum state is described by a unitary matrix $U \in \C^{2^n \times 2^n}$; i.e., a matrix such that $U^\dagger U= UU^\dagger =I$. 
These matrices preserve norms, mapping quantum states to other valid quantum states.
Any quantum algorithm, excluding measurements, corresponds to a unitary matrix. 
These matrices can be decomposed in one and two-qubits quantum gates, the basic elements of quantum circuits. 

Evolutions can be combined -- multiplications and tensor products of unitary matrices are still unitary. 
In practice, the reader might find it convenient to think of multiplications of unitary matrices as circuits applied to the same qubits (in series) and tensor products as circuits applied to different qubits (in parallel).
The evolution of quantum states is a unitary operation that requires quantum gates to be reversible.
Classical non-reversible gates (such as the $\mathtt{AND}$) have reversible counterparts ($\mathtt{CCNOT}$), implemented at the cost of introducing some auxiliary qubits.

\noindent b) Reading data from quantum states is not as immediate as accessing a memory location on a classical computer. 
While quantum states can be in a superposition of the computational basis, quantum mechanics doesn’t allow us to retrieve all the information at once from these superpositions.
\emph{Measurements} are modeled through a set of measurement operators $\{M_m\}_{m=0}^{2^n-1}$ such that $\sum_{m=0}^{2^n-1} M_m^\dagger M_m = 1$, with each $M_m \in \C^{2^n}$.
The probability that an outcome $m$ occurs for a state $\ket{\varphi}$ is given by $p(m) = \bra{\varphi}M_m^\dagger M_m\ket{\varphi}$, and after the $m^\thh$ outcomes is measured, the state \emph{collapses} on a new state $\ket{\varphi'} = \frac{M_m\ket{\varphi}}{\sqrt{\bra{\varphi}M_m^\dagger M_m\ket{\varphi}}}$.
In this work, we can restrict the measurement operators to $\{\ketbra{i}\}_{i=0}^{2^n-1}$. 
In this case, the resulting state of the quantum register is a vector of the computational basis.
In practice, if we measure a register (or a portion of it), we can only see one bit-string corresponding to one computational basis state that the register decomposes onto  (a sequence of 0s and 1s, like in a classical register). 

After the measurement, the register (or the qubits read) will lose the superposition and collapse on the measured state without providing further information on the amplitudes.
The process of reconstructing the amplitudes of a quantum state, with respect to a given basis, requires statistics over several measurements and is called \emph{tomography of pure states}.
The reversibility of quantum mechanics prevents a generic copying algorithm from existing (\emph{no-cloning} theorem). 
To perform tomography, one must re-create the quantum state by repeating the algorithms from scratch multiple times and sample (Theorem~\ref{thm:tomography}).


\section{Algorithms and subroutines implemented}
\label{app:math}

In this section, we detail the algorithms that we simulated to carry on the case study.
Starting from the input procedures, we precisely define the meaning of quantum access to a matrix.

\begin{definition} [Quantum access to a matrix \cite{kerenidis2020quantumGRADIENT}]
\label{def:quantum_access}
    We have quantum access to a matrix $A \in \mathbb{R}^{n \times m}$, if there exists a data structure that allows performing the mappings $\ket{i} \ket{0} \mapsto \ket{i}\ket{a_{i, \cdot}} = \ket{i}\frac{1}{\norm{a_{i,\cdot}}}\sum_j^m a_{ij}\ket{j}$, for all $i$, and $\ket{0} \mapsto \frac{1}{\norm{A}_F}\sum_i^n \norm{a_{i,\cdot}} \ket{i}$ in time $\widetilde{O}(1)$.
\end{definition}

This definition directly extends to vectors, which are special kind of matrices.
We say to have quantum access to a vector of size $m$ if we can implement the mapping
\begin{align}
    U_x\ket{0} = \frac{1}{\|x\|}\sum_{i \in m} x_i \ket{i}
\end{align}
its controlled version, and inverse, in time $\widetilde{O}(1)$.

Having access to a quantum random access memory, it is possible to efficiently create quantum access to matrices and vectors. 
This requires preprocessing the input data and storing a tree data structure, also known as KP-trees, after the authors of this procedure, in the QRAM. 
We state the main result and invite the interested reader to check the details in Kerenidis et al.~\cite{kerenidis2016quantumREC, kerenidis2020quantumGRADIENT}.

\begin{theorem}[Implementing quantum operators using an efficient data structure {\cite{kerenidis2016quantumREC}}]
\label{thm:efficient data structure}
    Let $A \in \R^{n\times m}$.
    There exists a data structure to store the matrix $A$ with the following properties:
    \begin{enumerate}
        \item The size of the data structure is $O(nm\log(nm))$\footnote{We use $\log(nm)$ instead of $\log^2(nm)$ because the extra $\log(nm)$ term in the original statement comes from the size of the system word.}.
        \item The time to update/store a new entry $(i,j,A_{ij})$ is $O(\log(nm))$.
        \item Provided coherent quantum access to this structure there exists quantum algorithms that implement the mappings of Def.~\ref{def:quantum_access} in time $O(\mathrm{polylog}(nm))$.
    \end{enumerate}
\end{theorem}

This definition of quantum access makes a data normalization parameter appear in many quantum algorithms. 
For a matrix $X$, we call this normalization parameter $\mu(X)$.
The smaller this parameter is, the more efficient the algorithms are.
The appendices of Kerenidis et al.~\cite{kerenidis2016quantumREC} describe how to implement the data structure for $\mu(X) = \norm{X}_F$. 
A subsequent manuscript by Kerenidis et al.~\cite{kerenidis2020quantumGRADIENT} describes how to obtain other values of $\mu$, which we describe in the following definition.

\begin{definition}[Memory parameter $\mu(X)$ \cite{kerenidis2020quantumGRADIENT}]
\label{def:mu}
Let $X \in \R^{n \times d}$ be a matrix. 
We define the parameter $\mu(X) = \min_{p \in [0,1]}(\norm{X}_F, \sqrt{s_{2p}(X)s_{2(1-p)}(X^T)})$, with $s_{q}(X) = \max_{i} \norm{X_{i,\cdot}}_q^q$, for $q \in [0,2]$.
\end{definition}

It is possible to probe the optimal $\mu$ during the data loading preprocessing step.
In practice, if the dataset is entirely available at loading time, one could measure $\mu$ for different values of $p \in [0,1]$ and compare it to the Frobenius norm, to know which data structure is more efficient.
The simulation of the quantum access routine is not required by our evaluation framework, while the estimation of the best $\mu$ is.

To retrieve data from quantum algorithms one has to consider some tomography procedures.
The algorithms that we consider leverage the following tomography routine.

\begin{theorem}[Tomography\cite{kerenidis2020quantumIP}]\label{thm:tomography}
Given access to the mapping $U_x\ket{0} \mapsto \ket{x}$ and its controlled version, for $\delta > 0$, there is an algorithm that produces an estimate $\overline x \in \mathbb{R}^m$ with $\|x\|_2 = 1$ with probability at least $1-1/\text{poly}(m)$ using $U$ $O(\frac{m\log m}{\delta^2})$ times such that $\| \ket{x} - \overline{x} \|_2 \leq \delta$ and using $U$ $O(\frac{\log(d)}{\delta^2})$ such that  $\| \ket{x} - \overline{x} \|_\infty \leq \delta$
\end{theorem}

Now that we discussed input and output routines, we move on and present the main building blocks of the quantum algorithms considered in this work: phase estimation, amplitude estimation, and distance/inner products estimation.
Each of this routine builds on the previous one.

\begin{theorem}[Phase estimation \cite{nielsen2010quantum}]\label{thm:phase-estimation}
Let $U$ be a unitary operator with eigenvectors $\ket{v_j}$ and eigenvalues $e^{i \theta_j}$ for $\theta_j \in [-\pi, \pi]$, i.e. we have $U\ket{v_j} = e^{i \theta_j}\ket{v_j}$ for $j \in [n]$. For a precision parameter $\epsilon > 0$, there exists a quantum algorithm that runs in time $O(\frac{T(U)\log(n)}{\epsilon})$ and with probability $1 - 1/poly(n)$ maps a state $\ket{\phi_i} = \sum_{j \in [n]} \alpha_j\ket{v_j}$ to the state $\sum_{j \in [n]} \alpha_j \ket{v_j}\ket{\bar{\theta}_j}$ such that $|\bar{\theta}_j - \theta_j| < \epsilon$ for all $j \in [n]$.
\end{theorem}

This procedure can be made consistent, in the sense that multiple runs of the same algorithm return a phase with the same error~\cite{ta2013inverting,kerenidis2020quantumgaussian}.
In our case study, we simulate this version of phase estimation, which is a consistent phase estimation.
We model the error using both the theory behind phase estimation and the procedure that makes it consistent.
Using phase estimation, we can build an amplitude estimation algorithm.

\begin{theorem}[Amplitude estimation\cite{brassard2002quantum}]\label{thm:amplitude estimation}
There is a quantum algorithm called amplitude estimation which takes as input one copy of a quantum state $\ket{\psi}$, a unitary transformation $U=2\ket{\psi}\bra{\psi}-I$, a unitary transformation $V=I-2P$ for some projector $P$, and an integer $t$. The algorithm outputs $\tilde{a}$, an estimate of $a=\braket{\psi|P|\psi}$, such that:
  $$|\tilde{a}-a| \leq 2\pi\frac{\sqrt{a(1-a)}}{t} + \frac{\pi^2}{t^2}$$
with probability at least $8/\pi^2$, using $U$ and $V$ $t$ times each. If $a=0$ then $\tilde{a}=0$ with certainty, and if $a=1$ and $t$ is even, then $\tilde{a}=1$ with certainty.
\end{theorem}

Using amplitude estimation on a modified Hadamard test circuit, it is possible to estimate distances and inner products.

\begin{theorem}\label{thm:innerproductestimation}
[Distance and Inner Products Estimation \cite{qmeans}] 
Assume for a matrix $V \in \mathbb{R}^{n \times d}$ and a matrix $C \in \mathbb{R}^{k \times d}$ that the following unitaries
$\ket{i}\ket{0} \mapsto \ket{i}\ket{v_i}$, and $\ket{j}\ket{0} \mapsto \ket{j}\ket{c_j}$
can be performed in time $T$ and the norms of the vectors are known. For any $\Delta > 0$ and $\epsilon>0$, there exists a quantum algorithm that computes:
$\ket{i}\ket{j}\ket{0}  \mapsto   \ket{i}\ket{j}\ket{\overline{d^2(v_i,c_j)}}$ where $|\overline{d^{2}(v_i,c_j)}-d^{2}(v_i,c_j)| \leqslant  \epsilon$ w.p. $\geq 1-2\Delta$ in time $\widetilde{O}\left(\frac{ \norm{v_i}\norm{c_j} T \log(1/\Delta)}{ \epsilon}\right)$.
\end{theorem}

Leveraging phase estimation, it is possible to implement one of the main ingredient of the quantum PCA model extraction pipeline of Figure~\ref{fig:quantum training}: the binary search for $\theta$.

\begin{theorem}[Quantum binary search for the singular value threshold \cite{qadra}] 
\label{Theorivisto:binarysearch} 
    Let there be quantum access to a matrix $\m{A} \in \R^{n \times m}$. 
    Let $\eta, \epsilon$ be precision parameters, and $\theta$ be a threshold for the smallest singular value to consider.
    Let $p \in [0,1]$ be the factor score ratios sum to retain.
    There exists a quantum algorithm that runs in time $\widetilde{O}\left(\frac{\mu(\m{A})\log(\mu(\m{A})/\epsilon)}{\epsilon\eta}\right)$ and outputs an estimate $\theta$ such that $\abs{p - \sum_{i: \overline{\sigma}_i \geq \theta} \lambda^{(i)}} \leq \eta$, where $\abs{\overline{\sigma}_i - \sigma_i} \leq \epsilon$, or detects whether such $\theta$ does not exists.
\end{theorem}

Similarly, with the aid of quantum tomography, it is possible to extract the top principal components and corresponding eigenvalues.

\begin{theorem} [Top-k singular vectors extraction {\cite{qadra}}] 
\label{TheoMio:top-k_sv_extraction}
    Let there be efficient quantum access to a matrix $\m{A} \in \R^{n \times m}$, with singular value decomposition $\m{A} = \sum_i^r \sigma_i \ve{u}_i \ve{v}_i^T$. Let $\delta > 0$ be a precision parameter for the singular vectors, $\epsilon>0$ a precision parameter for the singular values, and  $\theta>0$ be a threshold such that $\m{A}$ has $k$ singular values greater than $\theta$. Define $p=\frac{\sum_{i: \overline{\sigma}_i \geq \theta} \sigma_i^2}{\sum_j^r \sigma_j^2}$. There exist quantum algorithms that estimate: The top $k$ left singular vectors $\ve{u}_i$ of $\m{A}$ with unit vectors $\overline{\ve{u}}_i$
        such that $\norm{\ve{u}_i-\overline{\ve{u}}_i}_2 \leq \delta$ with probability at least $1-1/poly(n)$, in time $\widetilde{O}\left(\frac{\norm{A}}{\theta}\frac{1}{\sqrt{p}}\frac{\mu(\m{A})}{\epsilon}\frac{kn}{\delta^2}\right)$; 
  The top $k$ singular values $\sigma_i$, factor scores $\lambda_i$, and factor score ratios $\lambda^{(i)}$ of $\m{A}$ to precision $\epsilon$, $2\epsilon\sqrt{\lambda_i}$, and $\epsilon\frac{\sigma_i}{\norm{A}^2_F}$ respectively, with probability at least $1 - 1/\text{poly}(m)$, in time $\widetilde{O}\left(\frac{\norm{A}}{\theta}\frac{1}{\sqrt{p}}\frac{\mu(\m{A})k}{\epsilon}\right)$ or during any of the two procedures above.
\end{theorem}

Finally, building on the distance and inner product estimation routine, one can use a quantum version of $k$-means.
\begin{theorem}[$q$-means \cite{qmeans}]\label{thm:q-means}
Assume to have quantum access to a data matrix $V \in \mathbb{R}^{n \times d}$. For $\delta >0$, the q-means algorithm with high probability outputs centroids  $\{\overline{\mu_j} \}_{j=1}^k$ that are $\delta$-close in $\ell_2$ norm to the centroids of the classical $k$-means algorithm in time 
$$\widetilde{O}\left(    k d \frac{\eta}{\delta^2}\kappa(V)(\mu(V) + k \frac{\eta}{\delta}) + k^2 \frac{\eta^{1.5}}{\delta^2} \kappa(V)\mu(V)
\right)$$ per iteration with $1\leq \norm{v_i}^2 \leq \eta$ and a number of iterations proportional to the classical algorithm.
\end{theorem}

The simulation code is available on github\footnote{\url{https://github.com/tommasofioravanti/sq-learn}} and more info can be found in Fioravanti~\cite{tommythesis}.

\section{Vector state tomography}
\label{sec:tomography_est}
\begin{figure}[!ht]
    \centering
    \includegraphics[width=0.35\textwidth]{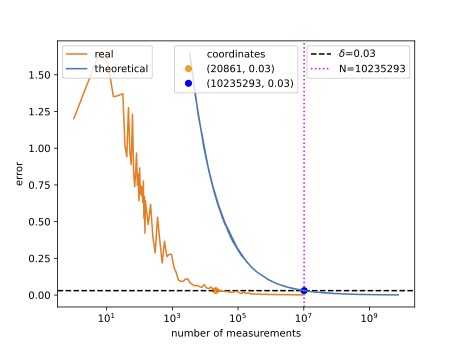}
    \caption{Theoretical bound (blue) on the number of measurements required for tomography and a numerical simulation (orange), on the first principal component of CIC-MalMem2022. The horizontal dashed line indicates the target $\ell_2$ error ($0.03$) of a vector of length $55$. The vertical dashed line represents the theoretical bound.}
    \label{fig:real_theory_a}
\end{figure}

The number of measurements, $N=\frac{36d\log d}{\delta^2}$ is the result of a probabilistic bound that  guarantees a success probability greater than $1 - 1/\mathrm{poly}(d)$.
Maintaining this failure probability low is crucial in applications where tomography has to be repeated many times on different vectors -- imagine its use in the prediction stage of a machine learning model -- as its repeated use will eventually lead to an estimate that exceeds the target error.
In this section, we exhibit a case where the number of samples $N=\frac{36d\log d}{\delta^2}$ is larger than the effective number of samples needed to obtain a certain accuracy: i.e., with a number of measurements considerably lower than $N$ we obtain an estimate $\overline{\boldsymbol{x}}$ with error $\delta$. 
We perform tomography over the first principal component of the dataset CIC-MalMem-2022\cite{carrier2022detecting}. 
The vector has size $d=55$. This experiment is reported in Figure~\ref{fig:real_theory_a}. 
The plot shows how many measurements ($x$-axis) are necessary to get a vector estimate with a specific error ($y$-axis), both following the theoretical bound (blue curve) and simulating the actual tomography procedure (orange curve).
\begin{figure}[!ht]
    \centering
    \includegraphics[width=0.35\textwidth]{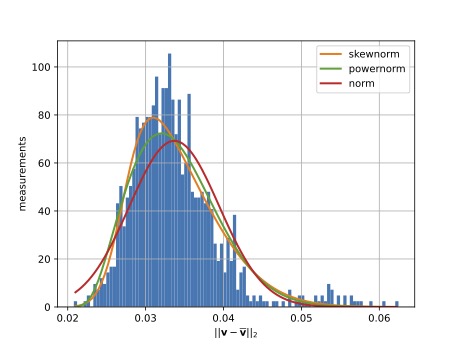}
    \caption{Distribution of error ($x$-axis) for tomography of the first principal component of CIC-MalMem2022 by repeating the experiment $1000$ times with $20,861$ samples. From this statistic with fit and plot three distributions (skew normal, power normal, and normal distribution).}
    \label{fig:real_theory_b}
\end{figure}
To get an estimate with error $0.03$ in $\ell_2$ distance, we need about $10^4$ measures instead of $\approx10^7$, as suggested by the theoretical bound. 
To corroborate this finding, in Figure~\ref{fig:real_theory_b}, we start from a fixed number of measurements $(20.861)$ (which is the number of measurement sufficient to reach error $0.03$ from Figure~\ref{fig:real_theory_a}), and we repeat the tomography for $1000$ times, plotting on the $y$-axis the frequency of the $\ell_2$ error observed. 
As we can see, the error is roughly centered around $0.03$.s

\begin{table}
\centering
\caption{Comparison for classical \textcircled{c} (PCA70) and quantum \textcircled{q} (QPCA70) principal components classifier with major and minor components over KDDCUP99.}
\label{table:kddcup pcc maj min}
\resizebox{\columnwidth}{!}{
\begin{tabular}{|c|cc|cc|cc|cc|} 
    \hline
$\alpha$ (\%) & \multicolumn{2}{c|}{Recall (\%)}
            & \multicolumn{2}{c|}{Precision (\%)}& \multicolumn{2}{c|}{F1-score (\%)}
            & \multicolumn{2}{c|}{Accuracy (\%)}
              \\
    \hline
      &   c  &   q  &   c  &   q &   c  &   q  &   c  &   q   \\
    \hline
1  &  98.62    & 98.68     &97.26     &96.99  &  97.94    & 97.83     &98.75    &98.68     \\
    
2&  98.66 &98.73       &95.96       &95.58 &  97.29 & 97.13       &98.35       &98.24                      \\
   
4   &98.80       &98.90       &91.61       &91.42  &95.07       &95.01       &96.91       &96.88                           \\
    
6   &98.84       &98.99       &88.88       &88.25     &93.60       &93.31       &95.78       &95.73                  \\

8   &98.93       &99.27       &86.11       &85.84   &92.08       &92.07       &94.88       &94.85           \\
   
10   &99.59       &99.87       &83.25       &83.14   &90.69       &90.74       &93.85       &93.87                 \\
\hline

\end{tabular}}
\end{table}

\section{Performance analysis with minor and major components over KDDCUP99}
\label{sec:pcckddmajorminor}
We corroborate the efficacy of the classifier discussed in Section~\ref{ssec:pcckdd} with another experiment. 
In this setting, we classify a sample as an attack if $(T_1>c_1 \textbf{ or } T_2>c_2)$, and as normal otherwise.
We maintain the same error parameters reported in Section~\ref{ssec:pcckdd} to estimate both major and minor components (QPCA70 with $10$ major and $7$ minor components, with respect to $10$ major and $6$ minor of the classical case). To extract the minor components, we set the threshold to $\theta=\sqrt{0.20}$ as a parameter in the quantum least-q singular vectors extraction (Theorem~\ref{TheoMio:top-k_sv_extraction}). 
With these error parameters, we extract a number of principal components very close to the classical model (QPCA70 with $10$ major and $7$ minor components, with respect to $10$ major and $6$ minor of the classical case). 
Also in this case, we match the classical performance, reported in Table~\ref{table:kddcup pcc maj min}. 
When using both major and minor components, the recall is higher than the model that used only major components (as reported in Table~\ref{table:q_exp1}), but precision is lower. 
This is expected as using two control summations in OR  (checking both $T_1$ and $T_2$, rather than $T_1$ only) leads to a higher chance of classifying an observation as an anomaly, resulting in an increase in false positives and a decrease in false negatives (hence improving recall). 

\section{Clustering with PCA-based dimensionality reduction and K-means over KDDCUP99}
\label{apx:q-means}

\begin{figure}
\centering
    \begin{subfigure}{0.49\columnwidth}
    \includegraphics[width=\columnwidth]{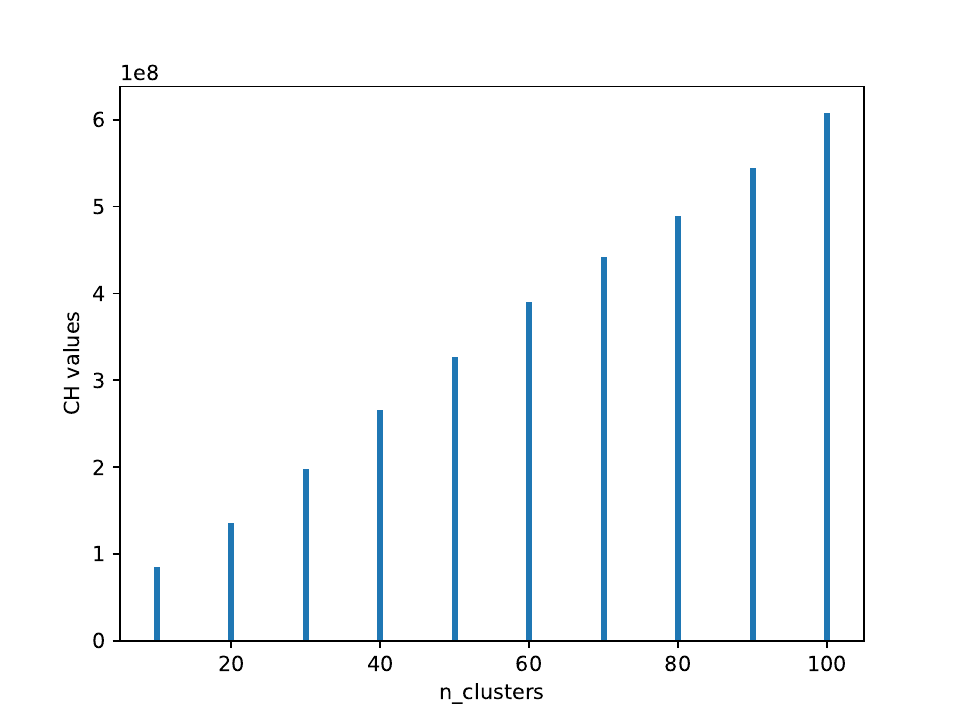}
    \caption[CH results]{Classical CH.}
    \label{fig:CH_classic}
    \end{subfigure}
    \begin{subfigure}{0.49\columnwidth}
    \includegraphics[width=\columnwidth]{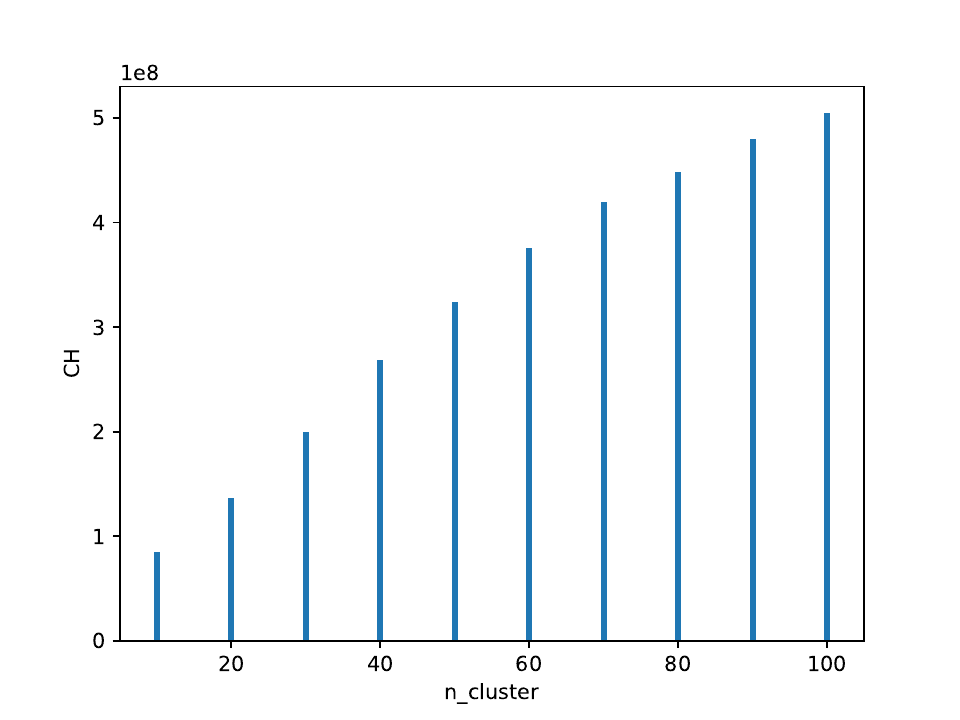}
    \caption[CH results]{Quantum CH.}
    \label{fig:CH_quantum}
    \end{subfigure}
    \caption{CH comparison between PCA and K-Means and q-PCA and q-Means, by varying the number of clusters $n_k$}
    \label{fig:CH_comparison}
\end{figure}

In this section, we demonstrate the feasibility of applying our methodology to the clustering task.
We perform K-Means after applying PCA-based dimensionality reduction and compare the classical and quantum algorithm versions. 
We evaluate the clustering quality of the classical and quantum algorithm versions by computing the \textit{Calinski-Harabasz (CH) index~\cite{scikit}} on the resulting clusters. 

For this experiment, we use the KDDCUP99 dataset, projecting all the data into a new 1-dimension PCA space and applying the K-Means clustering. We vary the number of clusters $n_k=[10,20,30,\dots,100]$ and compute the CH index. 
Since the PCA model retains only the first principal component, we classically compute the percentage of variance retained by the first principal component, which corresponds to $\approx0.6$. We use this value as input parameter $p=0.6$ into the quantum binary search, with $\epsilon_{\theta}=\epsilon=5$ and $\eta=0.1$. Then, we extract the top-k components with error $\delta=0.1$. After retrieving a classical description of the first principal component, we use it to project the data into the q-PCA feature space.
We apply the q-means algorithm over these 1-dimensional data with error $\delta=0.0005$ with  $n_k$ clusters. Once obtained a classical description of the clustering, we compute the CH score.
As shown In Figure~\ref{fig:CH_classic}, we match the classical CH index values.

\mypar{Further information on experiments}\label{ssec:furtherexp}

For more detail on the wrapper approach and the validation process we used for hyperparameter optimization we refer the interested reader to~\cite{wrapperapproach, optuna_2019}. 
Our hyperparameter tuning in Section~\ref{ssec:pca_loss} found that the number of quantiles that maximizes the F1 score is $751$.

\end{document}